\documentclass[journal]{IEEEtran}

\usepackage[noadjust]{cite}

\ifCLASSINFOpdf
   \usepackage[pdftex]{graphicx}
   \graphicspath{{img/}}
   \DeclareGraphicsExtensions{.pdf,.jpeg,.png}
\else
\fi
\usepackage{soul, color}
\usepackage{colortbl}
\usepackage{multirow}

\usepackage{enumitem}

\usepackage{lipsum}

\usepackage{amsmath}
\usepackage{amssymb}
\usepackage{amsfonts, bm}
\usepackage{steinmetz}
\usepackage{dsfont}
\usepackage[colorlinks=true]{hyperref}
\hypersetup{
    colorlinks,
    citecolor=blue,
    linkcolor=blue,
    urlcolor=blue
}

\usepackage{algorithmic}
\usepackage{algorithm}

\usepackage[caption=false, font=footnotesize]{subfig}

\usepackage{nomencl}
\makenomenclature

\usepackage{etoolbox}

\renewcommand\nomgroup[1]{%
  \item[\bfseries
  \ifstrequal{#1}{A}{General Math and Indices}{%
  \ifstrequal{#1}{B}{Grid Control Related}{%
  \ifstrequal{#1}{C}{RL Related}{}}}%
]}

\makeatletter
\def\ps@IEEEtitlepagestyle{
  \def\@oddfoot{\mycopyrightnotice}
  \def\@evenfoot{}
}
\def\mycopyrightnotice{
  {\scriptsize
  \begin{minipage}{\textwidth}
  \copyright 2022 IEEE. Personal use of this material is \href{https://journals.ieeeauthorcenter.ieee.org/become-an-ieee-journal-author/publishing-ethics/guidelines-and-policies/post-publication-policies/\#accepted}{permitted}.  Permission from IEEE must be obtained for all other uses, in any current or future media, including reprinting/republishing this material for advertising or promotional purposes, creating new collective works, for resale or redistribution to servers or lists, or reuse of any copyrighted component of this work in other works.
  \end{minipage}
  }
}

\usepackage{titlesec}

\titlespacing\section{0pt}{6pt plus 4pt minus 2pt}{0pt plus 2pt minus 2pt}
\titlespacing\subsection{0pt}{3pt plus 2pt minus 2pt}{0pt plus 2pt minus 2pt}
\titlespacing\subsubsection{0pt}{2pt plus 2pt minus 2pt}{0pt plus 2pt minus 2pt}

\usepackage{ctable}

\usepackage{titlesec}
\usepackage{amsmath,amssymb,amsthm,epsfig,color,empheq,graphicx}
\newtheorem{proposition}{Proposition}

\DeclareMathOperator*{\argmax}{\arg\!\max}

\begin{document}

\bstctlcite{IEEEexample:BSTcontrol}
%
\title{Curriculum-based Reinforcement Learning for Distribution System Critical Load Restoration}

\author{Xiangyu~Zhang, Abinet~Tesfaye~Eseye, Bernard~Knueven, Weijia~Liu, Matthew~Reynolds and Wesley Jones 

\thanks{\color{red}\textbf{Update:} \color{black}The curriculum learning software implementation of this paper is now open source at \url{https://github.com/NREL/rlc4clr}.}%
\thanks{This work was authored by the National Renewable Energy Laboratory, operated by Alliance for Sustainable Energy, LLC, for the U.S. Department of Energy (DOE) under Contract No. DE-AC36-08GO28308. Funding provided by the U.S. Department of Energy Office of Electricity (OE) Advanced Grid Modeling (AGM) Program. The views expressed in the article do not necessarily represent the views of the DOE or the U.S. Government. The U.S. Government retains and the publisher, by accepting the article for publication, acknowledges that the U.S. Government retains a nonexclusive, paid-up, irrevocable, worldwide license to publish or reproduce the published form of this work, or allow others to do so, for U.S. Government purposes.}%
\thanks{This research was performed using computational resources sponsored by the Department of Energy's Office of Energy Efficiency and Renewable Energy and located at the National Renewable Energy Laboratory.}
}

%
%

\markboth{T\MakeLowercase{his article has been accepted for publication in} IEEE T\MakeLowercase{ransactions on} P\MakeLowercase{ower} S\MakeLowercase{ystems}. DOI \href{https://doi.org/10.1109/TPWRS.2022.3209919}{10.1109/TPWRS.2022.3209919}}%
{Shell \MakeLowercase{\textit{et al.}}: Bare Demo of IEEEtran.cls for IEEE Journals}
%



\maketitle

\begin{abstract}
This paper focuses on the critical load restoration problem in distribution systems following major outages. To provide fast online response and optimal sequential decision-making support, a reinforcement learning (RL) based approach is proposed to optimize the restoration. Due to the complexities stemming from the large policy search space, renewable uncertainty, and nonlinearity in a complex grid control problem, directly applying RL algorithms to train a satisfactory policy requires extensive tuning to be successful. To address this challenge, this paper leverages the curriculum learning (CL) technique to design a training curriculum involving a simpler steppingstone problem that guides the RL agent to learn to solve the original hard problem in a progressive and more effective manner. We demonstrate that compared with direct learning, CL facilitates controller training to achieve better performance. To study realistic scenarios where renewable forecasts used for decision-making are in general imperfect, the experiments compare the trained RL controllers against two model predictive controllers (MPCs) using renewable forecasts with different error levels and observe how these controllers can hedge against the uncertainty. Results show that RL controllers are less susceptible to forecast errors than the baseline MPCs and can provide a more reliable restoration process.
\end{abstract}

\begin{IEEEkeywords}
Reinforcement learning, curriculum learning, critical load restoration, distribution system, grid resilience.
\end{IEEEkeywords}

%
\IEEEpeerreviewmaketitle







\section{Introduction}  \label{sec-introduction}

\IEEEPARstart{W}{ith} the increasing occurrence of extreme weather events, building a resilient distribution system that can withstand the impact of these events and mitigate the consequences is of significant importance \cite{panteli2015influence}. Following an outage, socially and economically critical loads need to be restored as soon as possible to provide basic societal needs and facilitate a successful overall recovery. As smart grid technologies develop, swift critical load restoration (CLR) is made feasible as a result of remotely controlled switches, which allow post-event system reconfiguration, and distributed energy resources (DERs), which enable local generation support. 
To conduct network reconfiguration, optimally dividing the system into self-sustained sections has been studied \cite{poudel2018critical, xu2016microgrids,li2014distribution, wang2019coordinating, chen2017multi}. In \cite{wang2019coordinating}, a two-stage method is proposed to first heuristically determine post-restoration topology and then solve a mixed-integer semidefinite optimization problem to determine the restored power flow. To further ensure the feasibility of intermediate restoration steps, a sequential service restoration framework is studied in \cite{chen2017multi}.
In addition to network reconfiguration, other studies, e.g., \cite{wang2018risk}, focus on DERs scheduling after the system is reconfigured. This is also the focus of this paper: we aim at studying how to optimally leverage DERs, both dispatchable and renewable, to maximize load pick-up or other resilience-related metrics.

Using renewables-based DERs in CLR makes the consideration of uncertainty critical as their generation might not be delivered as predicted. The following inexhaustive summary reviews some general approaches to hedge against the uncertainty used in load restoration literature. First, \textit{scenario-based} stochastic programming methods are studied, aiming at optimizing the expected performance under generated scenarios \cite{wang2015self, amin2016micro, yao2020rolling}. Second, \textit{robust optimization} handles uncertainty by ensuring the solution is feasible for the worst-case scenario. It is discussed in \cite{liu2020bi} for a bi-level service restoration problem considering the uncertainties of DERs and loads. Chen \textit{et al.} \cite{chen2016robust} develop a robust restoration method and introduces an uncertainty budget technique to adjust the conservativeness of the solution. Third, \textit{chance-constrained methods} that use probabilistic constraints to enforce the satisfaction of system constraints within some probability thresholds; see \cite{wang2018risk, gao2016resilience} for examples. Fourth, \textit{model predictive control (MPC)}, which re-optimizes the problem at every control interval, can leverage the latest forecasts with more accurate predictions to re-align the system with the optimal trajectory. Liu \textit{et al.} \cite{liu2020collaborative} propose using MPC for coordinating multiple types of DERs in distribution system restoration, and it is utilized in \cite{zhao2018model} to optimize generators' start-up for system restoration. 
Other MPC based restoration examples are \cite{yao2020rolling, eseye2021resilient}.

In addition to optimization-based methods, reinforcement learning (RL), which has proven effective in several sequential decision-making applications \cite{mnih2013playing, johannink2019residual}, represents a potential alternative to tackle the CLR problem with the following merits:
\begin{enumerate}
    \item RL methods train a control policy prior to online control and do not require intensive on-demand computation during the restoration process. This allows RL to provide restoration plan for a longer control horizon and a shorter control interval, i.e., problems that cannot be efficiently solved by optimization between control intervals. Once the policy is trained, RL can provide fast real time response as a result of the simplicity of policy evaluation.
    \item RL agents can learn from nonlinear power flow models and thus avoid losing control accuracy due to model simplification.
    \item RL can directly learn from historical data and conduct an end-to-end uncertainty management, eliminating the need to estimate uncertainty distributions and/or generate scenarios for stochastic programming.
\end{enumerate}
Motivated by these RL merits, Bedoya \textit{et al.} \cite{bedoya2021distribution} solve a CLR problem considering the asynchronous data arrival using deep Q-network (DQN) and \cite{zhao2021learning} proposed an approach combining a graph convolutional network with a DQN to conduct sequential system restoration. Unlike these papers, we focus on investigating RL's capability to manage uncertainty in CLR when provided with imperfect renewable forecasts.

However, because of the data-intensive nature of RL and the non-convexity of deep RL policy searching, training a performant controller to solve a real-world grid control problem is difficult. In practice, leveraging special techniques to facilitate the training process can be very beneficial. As demonstrated in \cite{lin1992reinforcement}, \textit{instructive training instances} and \textit{hierarchical learning} can be used to speedup learning. Later, Bengio \textit{et al.} \cite{bengio2009curriculum} further formalize this as \textit{curriculum learning (CL)}, which breaks down a complicated task into a series of tasks with gradually increased complexity, and introduces them sequentially to an RL agent to learn to tackle the original hard problem in a ``divide-and-conquer'' manner; see \cite{narvekar2020curriculum} for a comprehensive review. As demonstrated in \cite{bengio2009curriculum}, CL can 1) accelerate policy convergence and 2) improve the quality of the obtained local-optimum in the non-convex setting. Inspired by this idea, we design a two-stage curriculum to facilitate the efficient and effective training of the RL agent to perform the CLR control.

In summary, the main contributions of this paper include: 1) A CL-based RL training approach is developed to facilitate the CLR controller's training, enabling convergence to a better control policy; 2) We demonstrate the RL controller's capability to conduct reliable load restoration using imperfect renewable forecasts while satisfying operational constraints; 3) To the best of our knowledge, this paper presents the first study that systematically compares RL controllers with MPC baselines with the focus on their uncertainty hedging capability under controlled error levels in an electric grid control problem. Our prior work~\cite{zhang2021restoring} demonstrated a preliminary version of leveraging CL for the CLR problem, \emph{without} considering renewable generation forecast errors and did not contain a comparison with optimization-based MPC baselines.

The remainder of the paper proceeds as follows: Section II introduces the CLR problem formulation and Section III discusses the RL formulation for this problem. Section IV presents the methodology to generate realistic imperfect renewable forecasts used in this investigation. The proposed curriculum learning design is discussed in Section V and case study results are presented in Section VI. Finally, Section VII provides some discussion and conclusions.

\section{Critical Load Restoration Problem} \label{sec-clr-problem}

Consider a multi-step prioritized CLR problem after a distribution system is islanded from the main grid due to an extreme event. The goal is to restore as many critical loads as possible in the outage duration, denoted by discrete steps $\mathcal{T}=\{1, 2, ..., T\}$, using only local DERs. Specifically, critical loads $i \in \mathcal{L}$ in the system are prioritized by $\zeta^i$, and $\mathbf{z} = [\zeta^1, \zeta^2, ..., \zeta^N]^\top \in \mathbb{R}^{N}$ is the vector of all load priority factors ($N=|\mathcal{L}|$\footnote{In this paper, $|\cdot|$ calculates the cardinality of a set or the length of a vector, $||\mathbf{x}||_2$ calculates the Euclidean norm of vector $\mathbf{x}$, diag$\{\mathbf{x}\}$ refers to a diagonal matrix with vector $\mathbf{x}$ as diagonal elements. Superscript ``+'' of a vector indicates $[x^1, ..., x^n]^+ = [(x^1)^+, ..., (x^n)^+]$ and $(x^i)^+ = \text{max}(0, x^i)$. $\mathbf{0}_{k}$/$\mathbf{1}_{k}$ are $k$-dimensional vectors with all zeros/ones.}). Available DERs include three types: renewable-based ($\mathcal{R}$), dispatchable fuel-based ($\mathcal{D}^{f}$) and dispatchable energy storage ($\mathcal{D}^{s}$). We denote all DERs as $\mathcal{G} = \mathcal{R} \bigcup \mathcal{D}^{f} \bigcup \mathcal{D}^{s}$. In the following problem formulation, some assumptions are made:

\noindent $\boldsymbol{[A.1]}$ This study only considers the uncertainty from the renewable generation. The demand of critical loads $\mathcal{L}$ ($\mathbf{p} = [p^{1}, p^{2}, ..., p^{N}]^\top \in \mathbb{R}^{N}$ and $\mathbf{q} = [q^{1}, q^{2}, ..., q^{N}]^\top \in \mathbb{R}^{N}$), in contrast, are assumed to be time-invariant over $\mathcal{T}$ \cite{wang2018risk, poudel2018critical}. We also think loads can be partially and continuously restored, e.g., restoring 83.72 kW out of 100 kW requested, assuming intelligent grid edge devices can help approximate such a finer level of load control via smart switches at customer end. 

\noindent $\boldsymbol{[A.2]}$ This paper considers only steady-state analysis regarding DERs dispatch and load pick-up scheduling in CLR, and we defer the study of frequency dynamics and stability during the restoration to future works; interested readers are referred to \cite{xu2016microgrids, zhang2021two} on involving dynamic analysis in a service restoration study. 

\noindent $\boldsymbol{[A.3]}$ We assume, at the beginning of the control horizon, i.e., $t=1$, a radial distribution network is already reconfigured and energized; all DERs are brought in synchronous by black-start capable ones and are ready to pick up loads. The network topology remains the same during the restoration. These assumptions are made to focus on the post-reconfiguration DERs scheduling problem described below.

\subsection{Objective}  \label{subsec-opt-obj}

At each control step $t \in \mathcal{T}$, the controller determines the active power set points and power factor angles for all DERs (i.e., $\mathbf{p}^\mathcal{G}_t \in \mathbb{R}^{|\mathcal{G}|}$ and $\bm{\alpha}^\mathcal{G}_t \in \mathbb{R}^{|\mathcal{G}|}$) and the amount of demand restored for all critical loads (i.e., $\mathbf{p}_t = [p_t^{1}, p_t^{2}, ..., p_t^{N}]^\top \in \mathbb{R}^{N}$ and $\mathbf{q}_t = [q_t^{1}, q_t^{2}, ..., q_t^{N}]^\top \in \mathbb{R}^{N}$), to \textit{maximize} the following control objective function:

\begin{equation} \label{eq-obj-func}
\sum_{t\in \mathcal{T}} (r_t^{\text{CLR}} + \vartheta_t),
\end{equation}
in which

\begin{equation}  \label{eq-clr-reward}
    r_t^{\text{CLR}} := \mathbf{z}^\top \mathbf{p}_t - \mathbf{z}^\top \text{diag}\{\bm{\epsilon}\}[\mathbf{p}_{t-1} - \mathbf{p}_{t}]^+
\end{equation}
represents the single step load restoration reward, and

\begin{equation}  \label{eq-vv-penalty}
\vartheta_t := -\lambda ||[\mathbf{v}_t - \overline{\mathbf{v}}]^+ + [\underline{\mathbf{v}} - \mathbf{v}_t]^+ ||_2^2
\end{equation}
shows the single step voltage violation penalty over all $N_b$ buses, where $\mathbf{v}_t \in \mathbb{R}^{N_b}$, $\underline{\mathbf{v}} = V^{\text{min}} \mathbf{1}_{N_b} \in \mathbb{R}^{N_b}$ and $\overline{\mathbf{v}} = V^{\text{max}} \mathbf{1}_{N_b} \in \mathbb{R}^{N_b}$ represent the voltage magnitude of all buses at time $t$, and the normal voltage limits (e.g., ANSI C.84.1 limits).
Specifically, in (\ref{eq-clr-reward}), the first term encourages load restoration, and the second term penalizes shedding previously restored load by factors of $\bm{\epsilon}=[\epsilon^1, ... , \epsilon^N]^\top \in \mathbb{R}^N$. Introducing this penalty is to facilitate a reliable and monotonic restoration and thus minimize the impact from the intermittent renewable generation. Grid operators can choose $\epsilon^i$ to control the strictness of the monotonic load restoration requirement since based on \eqref{eq-clr-reward} the controller should only restore load $i$ if it can be sustained for the next $\epsilon^i+1$ steps to keep the overall reward positive. Meanwhile, the added penalty also introduces strong temporal-dependency among control steps and makes the problem more challenging. A large positive value coefficient $\lambda$ in (\ref{eq-vv-penalty}) encourages bus voltages to be within limits. Note, voltage bounds are enforced as a penalty term because voltage values are system-controlled outcomes and cannot be directly constrained by RL formalism. In summary, maximizing (\ref{eq-obj-func}) is equivalent to maximizing the area below the \textit{resilience curve} \cite{panteli2015grid} and thus enhancing the grid resilience \cite{wang2018risk, gao2016resilience}. 

\subsection{Operational Constraints}  \label{subsec-clr-constraints}

During the CLR process, the following operation constraints should be satisfied for all $t \in \mathcal{T}$:

1) \textit{Fuel-based DERs}: For $\forall g \in \mathcal{D}^{f}$, active power, power factor angle, and total generation energy follow:

\begin{equation}  \label{eq-fuel-der-constraints}
    p_t^{g} \in [\underline{p^{g}}, \overline{p^{g}}],\quad
    \alpha_t^g \in [\underline{\alpha^g}, \overline{\alpha^g}],\quad
    \sum_{t \in \mathcal{T}} p_t^{g} \cdot \tau \leq E^{g}
\end{equation}
in which $\tau$ is the control interval (unit: \textit{hour}), $E^{g}$ represents the known maximum energy limit (e.g., fuel reserve) and $\alpha_t^g$ is the operating power factor angle ($\alpha_t^g = \text{arctan}(q_t^{g}/p_t^{g})$). Ramping rate limits are not considered here, but can be considered if using $\Delta p_t^{g} \in [\underline{\Delta p^{g}}, \overline{\Delta p^{g}}]$ as decision variables.

2) \textit{Energy Storage (ES)}: This study assumes the energy storage can charge/discharge at any power within the feasible range. The state of charge (SOC) feasibility, charging/discharging state transition, initial storage and power factor angle, for $\forall \theta \in \mathcal{D}^{s}$, are constrained by:

\begin{align}  \label{eq-storage-constraints}
\begin{split}
p_t^{\theta} \in [-p^{\theta, \text{ch}}, p^{\theta, \text{dis}}] &, \quad
S_{t+1}^{\theta} = S_{t}^{\theta} - \eta_t \cdot p_t^{\theta} \cdot \tau\\
S_{t}^{\theta} \in [\underline{S^{\theta}}, \overline{S^{\theta}}] &, \quad
S_0^{\theta} = s_0, \quad
\alpha_t^{\theta} \in [\underline{\alpha^{\theta}}, \overline{\alpha^{\theta}}],
\end{split}
\end{align}
in which $\eta_t$ is the efficiency factor and $\eta_t = 1 / \eta^{dis}$ for times $t$ when battery is discharging ($p_t^{\theta} > 0$) and $\eta_t = \eta^{ch}$ for times $t$ when the battery is charging ($p_t^{\theta} < 0$). This causes non-linearity, but can be avoided in implementation by introducing additional binary charging/discharging status variables. $S_t^{\theta}$ and $s_0$ are the current and initial SOC, respectively.

3) \textit{Renewable DERs}: Renewable generation is limited by available natural resources and will be maximally utilized during restoration. The power factor angle should satisfy the inverter's limits. Specifically, for $\forall r \in \mathcal{R}$, there are:

\begin{equation}  \label{eq-renewable-constraints}
    p_t^{r} = \hat{p}_t^{r}, \quad
    \alpha_t^r \in [\underline{\alpha^r}, \overline{\alpha^r}].
\end{equation}
$\hat{p}_t^{r}$ is the time-variant renewable generation determined by the natural resource. Note $\hat{\cdot}$ is the symbol for forecasts since $\hat{p}_t^{r}\ (\forall t \in \mathcal{T})$ appear in this multi-step scheduling problem as predicted values and are the sources of uncertainty. 

4) \textit{Loads}: The load pick-up decision should satisfy:

\begin{equation}  \label{eq-load-constraints}
    \mathbf{0}_{N} \leq \mathbf{p}_t \leq \mathbf{p}, \quad
    \mathbf{0}_{N} \leq \mathbf{q}_t \leq \mathbf{q}, \quad
    p_t^i/q_t^i = p^i/q^i,
\end{equation}
which follows the aforementioned assumption \textbf{A.1}.

5) \textit{Power Balance and Power Flow Constraints}: Since the distribution system is islanded, the balance between loads and DERs generation should be satisfied. Additionally, the electrical relationship among related values is enforced by a power flow model $f$. Together, these lead to:

\begin{subequations}  \label{eq-power-flow-constraints}
\begin{align}
    &\mathbf{1}_N^{\top} \mathbf{p}_t = \mathbf{1}_{|\mathcal{G}|}^{\top} \mathbf{p}_t^{\mathcal{G}}, \ \mathbf{1}_N^{\top} \mathbf{q}_t = \mathbf{1}_{|\mathcal{G}|}^{\top} (\mathbf{p}_t^{\mathcal{G}} \odot tan(\bm{\alpha}^\mathcal{G}_t)), \\
    &\mathbf{v}_t = f (\mathbf{p}_t, \mathbf{q}_t, \mathbf{p}^\mathcal{G}_t, \bm{\alpha}^\mathcal{G}_t)
\end{align}
\end{subequations}
where $\odot$ represents the Hadamard product of two vectors and $tan([x^1, ..., x^n]^\top)=[tan(x^1), ..., tan(x^n)]^\top$. The general representation of power flow $f$ calculates bus voltages $\mathbf{v}_t$ from system power injection and $\mathbf{v}_t$ will then be used in \eqref{eq-vv-penalty} to evaluate $\vartheta_t$. Specifically, for the RL implementation, the distribution system simulator OpenDSS is used to instantiate $f$; while in optimization-based baseline controllers, a linear power flow model, LinDistFlow \cite{gan2014convex}, with the same parameters used for simulation, is utilized.

\subsection{Optimal Control Formulation}

Combining objective and constraints, mathematically, the optimal control problem focused in this study is given by:

\begin{equation}\tag{\bf{OCP}} \label{eq-opt-formulation}
\begin{aligned}
& \underset{\mathbf{p}_t, \mathbf{q}_t, \mathbf{p}^\mathcal{G}_t, \bm{\alpha}^\mathcal{G}_t, \forall t \in \mathcal{T}}{\operatorname{maximize}} 
& & (\ref{eq-obj-func}) \\
& \underset{\forall t \in \mathcal{T}}{\text{subject to}}
& & (\ref{eq-fuel-der-constraints}) - (\ref{eq-power-flow-constraints}),
\end{aligned}
\end{equation}
which is  formulated as a mixed-integer linear programming (MILP) problem. Though we formulate the CLR problem as an optimization problem here \textit{for better understanding}, in later discussion, RL is utilized to solve it. However, the RL controller's performance is compared against MPC baselines utilizing the MILP formulation of (\ref{eq-opt-formulation}). Specifically, the two baseline MPC controllers considered are:

1) \textbf{NR-MPC} \textit{(No Reserve MPC)}: This controller repeatedly solves (\ref{eq-opt-formulation}) sequentially in a receding horizon manner with updated renewable forecasts at each step, i.e., $\hat{p}_t^r, \forall r\in \mathcal{R}$ in (\ref{eq-renewable-constraints}). Because of the step-wise re-optimizing, uncertainty in renewable generation can be partially addressed.

2) \textbf{RC-MPC} \textit{(Reserve Considered MPC)}: RC-MPC \cite{eseye2021resilient} is a more robust version of MPC and it solves a modified problem similar to (\ref{eq-opt-formulation}), as detailed in Appendix \ref{appen-rc-mpc-formulation}. Specifically, the objective function is augmented with a penalty term which leads to an explicit consideration of system generation reserve to hedge against the error in renewable forecasts.

\section{The Reinforcement Learning Formulation} \label{sec-rl-controller}
This section defines the state, action and reward in the RL Markov Decision Process (MDP) formulation for solving (\ref{eq-opt-formulation}). 
RL preliminaries are not presented; interested readers are referred to \cite{sutton2018reinforcement}. 

\subsection{\textit{State}} \label{subsec-state}
State $\mathbf{s}_t$ is the input of an RL controller and is used for decision-making at each control step $t \in \mathcal{T}$. It contains information reflecting the system status, and we define $\mathbf{s}_t$ as:

\begin{equation}  \label{eq-state-definition}
\mathbf{s}_t := [(\hat{\mathbf{p}}_{t}^{\mathcal{R}})^\top, (\widetilde{\mathbf{p}}_{t-1})^\top, (\mathbf{S}_{t}^{\theta})^\top, (\mathring{\mathbf{E}}_{t}^{\mu})^\top, t, (\Phi_t)^\top]^\top \in \mathcal{S},
\end{equation}
where $\mathcal{S}$ is the RL state space.

First component of $\mathbf{s}_t$ is the generation forecasts for all renewables-based DERs $\mathcal{R}$, which are in practice provided by the grid operator's forecasting modules: 
\begin{subequations}  \label{eq-renewable-forecasts}
\begin{align}
    & \mathbf{\hat{p}}_{t}^\mathcal{R} := [(\hat{\mathbf{p}}_{t}^1)^\top, ..., (\hat{\mathbf{p}}_{t}^{|\mathcal{R}|})^\top]^\top \in \mathbb{R}^{k|\mathcal{R}|/\tau}, \label{eq-all-renewable-forecast-vector} \\
    & \mathbf{\hat{p}}_{t}^r := [p_t^r, \hat{p}_{t+1|t}^r, ..., \hat{p}_{t+(k/\tau) - 1|t}^r]^\top \in \mathbb{R}^{k/\tau}.  \label{eq-single-renewable-forecast-vector}
\end{align}
\end{subequations}
Here, $\hat{p}_{t+x|t}^r$ is the generation prediction for step $t+x$ of DER $r \in \mathcal{R}$, and the ``$\cdot|t$'' in the subscript indicates the forecast is made/updated at step $t$. Because $\tau$ represents control interval length in hours (see \eqref{eq-fuel-der-constraints} in Section \ref{subsec-clr-constraints}), the dimension $k/\tau$ represents the number of forecast data points for the $k$-hour look-ahead period (We choose $\tau$ that makes $1/\tau$ an integer). E.g., for $\tau=1/12$ and $k=2$, i.e., 5-min control interval and two-hour look-ahead, there is $|\mathbf{\hat{p}}_{t}^r|=24$. 

Second, $\widetilde{\mathbf{p}}_{t-1}:=\text{diag}\{\mathbf{p}\}^{-1}\mathbf{p}_{t-1} \in \mathbb{R}^N$ is used to reflect the fractional load restoration level at the previous step. We assume all loads are initially at 0 kW, i.e., $\widetilde{\mathbf{p}}_0=\mathbf{0}_N$.

Third, the remaining SOC and fuel for all the ES systems and fuel-based DERs, which imply the remaining load supporting capability from dispatchable DERs: $\mathbf{S}_t^{\theta} \in \mathbb{R}^{|\mathcal{D}^{s}|}$ and $\mathring{\mathbf{E}}_t^{\mu} \in \mathbb{R}^{|\mathcal{D}^{f}|}$. 

Lastly, some auxiliary variables are included: Current step index $t$ is included to inform the restoration progress and $\Phi_t = [sin(\frac{\text{hour}(t) * 2\pi}{24}), cos(\frac{\text{hour}(t) * 2\pi}{24})]^\top$ is the trigonometric encoding of the time of the day.

Specifically, regarding the renewable forecasts component $\mathbf{\hat{p}}_{t}^\mathcal{R}$, it is worth to point out the following:

1) Renewable generation forecasts are commonly used for grid control problems \cite{shen2020distributed}, so it is justifiable to include them in $\mathbf{s}_t$ to inform the future generation potential. Considering multi-step forecasts $\mathbf{\hat{p}}_{t}^r$, instead of using only current measurement, e.g., $p_t^r$, allows the controller to better capture the trend of the environment \cite{wei2017deep}.

2) Renewable generation forecasts are, in reality, inaccurate. Based on imperfect or even sometimes misleading forecasts $\mathbf{\hat{p}}_{t}^\mathcal{R}$, RL controller's performance is inevitably impacted; the same for MPC that solves \eqref{eq-opt-formulation}, inaccurate forecasts used in \eqref{eq-renewable-constraints} lead to erroneous formulation and thus suboptimal solution. Therefore, one emphasis in this study is to investigate and compare controllers for their robustness/performance deterioration under different error levels, low to high, in $\mathbf{\hat{p}}_{t}^\mathcal{R}$. Section \ref{sec-forecasts} will discuss how to study this problem in a controlled manner using synthetic forecasts.

3) The choice of the look-ahead length $k$ can also affect the controller performance: a larger $k$ includes more forecasts but makes the RL controller harder to train due to the enlarged $\mathbf{s}_t$. We will study this in Section \ref{sec-case-study}. 

4) The auxiliary variables $\Phi_t$, which tells the controller the time of day, are helpful to evaluate the solar generation resources if $k$ is small: e.g., at 10PM or 4AM, though $\mathbf{\hat{p}}_{t}^r=\mathbf{0}_{k/\tau}$ for both cases if $k=2$, the time encoding can help controller learn that in the second case the PV generation is more likely to be available soon as it is in early morning.

\subsection{\textit{Action}} Action $\mathbf{a}_t$ determines the control action values at each step $t\in\mathcal{T}$. Corresponding to (\ref{eq-opt-formulation}), the decision should include: i) load amount to be restored and ii) generation set points for DERs. So, the action has the following format:

\begin{equation} \label{eq-action-definition}
    \mathbf{a}_t := [(\mathbf{p}_t)^\top, (\mathbf{H}_p\mathbf{p}^\mathcal{G}_t)^\top, (\mathbf{H}_{\alpha}\bm{\alpha}^\mathcal{G}_t)^\top]^\top \in \mathcal{A},
\end{equation}
while the specific values at each step is determined by the RL policy, i.e., $\mathbf{a}_t=\pi(\mathbf{s}_t)$. By choosing how much load to restore and how to utilize the available generation resources in the dispatchable DERs (i.e., fuel or stored energy), $\mathbf{a}_t$ directly affects $\widetilde{\mathbf{p}}_{t}$, $\mathbf{S}_{t+1}^{\theta}$ and $\mathring{\mathbf{E}}_{t+1}^{\mu}$ in $\mathbf{s}_{t+1}$, and over the control horizon, the RL agent can determine a restoration trajectory according to its policy. In \eqref{eq-action-definition}, note that the load reactive power $\mathbf{q}_t$ is absent, this is because of $q_t^i = p_t^i \cdot (q^i / p^i), \forall i \in \mathcal{L}$ (recall Assumption \textbf{A.1} and (\ref{eq-load-constraints}) in Section \ref{sec-clr-problem}). In addition, since the distribution system is islanded, there should be a balance between loads and generation in the system, e.g., $\mathbf{1}_N^{\top} \mathbf{p}_t = \mathbf{1}_{|\mathcal{G}|}^{\top} \mathbf{p}_t^{\mathcal{G}}$ for active power. So, to facilitate this (not enforce though, see Section \ref{sec-case-study} for details), two selection matrices $\mathbf{H}_p\in\mathbb{R}^{(|\mathcal{D}^f| + |\mathcal{D}^s| -1)\times |\mathcal{G}|}$ and $\mathbf{H}_{\alpha}\in\mathbb{R}^{(|\mathcal{G}|-1)\times |\mathcal{G}|}$ are introduced, indicating that the active power outputs of dispatchable DERs and reactive power outputs of all DERs are being controlled. The ``$-1$'' in the dimension above is to leave one DER's power output out for power balancing purpose. So the unselected DER's power output is naturally determined by $\mathbf{1}_N^{\top} \mathbf{p}_t - \mathbf{1}^{\top}_{|\mathcal{D}^f| + |\mathcal{D}^s| - 1} \mathbf{H}_p \mathbf{p}_t^{\mathcal{G}} - \mathbf{1}_{|\mathcal{R}|}^{\top} \mathbf{p}^{\mathcal{R}}_t$, where $\mathbf{p}^{\mathcal{R}}_t$ is the renewable generation at step $t$. 

In summary, $\mathbf{a}_t$ determines load pickup amount for all loads and active/reactive power outputs for selected DERs at each step $t$.

\subsection{\textit{Reward}} Reward $r_t$ is an evaluation of the control $\mathbf{a}_t$ based on $\mathbf{s}_t$ in the form of a scalar, i.e., $r_t = R(\mathbf{s}_t, \mathbf{a}_t)$. Corresponding to (\ref{eq-obj-func}), the reward is straightforwardly defined as $r_t = r_t^{\text{CLR}} + \vartheta_t$, and it is calculated based on the simulation results at Step $t$. It's worth to point out that, $r_t$ is determined not only by the immediate action $\mathbf{a}_t$, but also implicitly by previous actions $\mathbf{a}_{t'} (t'<t)$ that drive the system to $\mathbf{s}_t$. So, the long-term impact of agent's action leads to a strong temporal-dependency in the formulated problem.

\subsection{Environment Interaction and RL Training}
By interacting with a simulation environment, an RL control policy, i.e., a state-action mapping relationship, 

\begin{equation} \label{eq-optimal-policy-def}
    \pi^*(\mathbf{a}_t|\mathbf{s}_t) = \argmax_{\pi} \mathbb{E}_{\mathbf{a}_t\sim\pi, \xi \sim \Xi^{\text{tr}}}\left(\sum_{t\in \mathcal{T}} r_t\right),
\end{equation}
is trained to guide the RL agent to make proper decisions at each step and to maximize the expected cumulative rewards over the control horizon $\mathcal{T}$. 
In this study, the policy is instantiated by a neural network that takes in 
$\mathbf{s}_t$ as an input and outputs the control actions $\mathbf{a}_t$. The RL agent then sends $\mathbf{a}_t$ to the learning environment for control implementation and obtains $\mathbf{s}_{t+1}$ and $r_t$ from the environment (see Section \ref{subsubsec-learning-env} for more details on the implementation).

The $\Xi^{\text{tr}}$ in \eqref{eq-optimal-policy-def} means the expectation is maximized over multiple training scenarios of renewable generation. Each scenario $\xi=\bigcup_{r\in\mathcal{R}} (\mathbf{p}^r_{1:T}, \mathbf{\hat{p}}_{1}^r, ..., \mathbf{\hat{p}}_{T}^r) \in \Xi^{\text{tr}}$ collects 1) an actual renewable generation profile $\mathbf{p}^r_{1:T}= [p_1^r, p_2^r, ..., p_{T}^r]^\top, \forall r\in\mathcal{R}$, which is \textit{unknown} to the RL agent and is used for simulation only; and 2) the renewable generation forecasts $\mathbf{\hat{p}}_{t}^r$, as an estimation of $\mathbf{p}^r_{t:T}$, generated by the grid operator's forecasting modules. The RL agent makes decision based on such forecasts since $\mathbf{\hat{p}}_{t}^r$ is part of $\mathbf{s}_t$ (see \eqref{eq-renewable-forecasts}). The reason to optimize the policy over multiple renewable generation scenarios is to ensure the RL controller can generalize well in unseen scenarios. To test that, the policy $\pi^*(\mathbf{a}_t|\mathbf{s}_t)$, once trained, is assessed using different testing scenarios $\Xi^{\text{ts}}$ for performance evaluation.

\section{Synthetic Renewable Generation Forecasts with Controlled Error Levels} \label{sec-forecasts}

To generate the training scenarios $\Xi^{\text{tr}}$, we use renewable generation historical data of $|\mathcal{R}|$ DERs derived from a public data set \cite{draxl2015wind}, denote as $\mathcal{P}^\mathcal{R}$, to sample realistic $\mathbf{p}^r_{1:T}$, i.e., $\mathbf{p}^r_{1:T} \sim \mathcal{P}^r \subset \mathcal{P}^\mathcal{R}$, for an outage duration. However, since the historical operational forecasts, i.e., $\mathbf{\hat{p}}_{t}^r$, are absent from the public data set, we propose a mechanism $\mathcal{Y}$ to generate \textit{synthetic forecasts} from $\mathbf{p}^r_{1:T}$ with a \textit{controlled error level $\epsilon$}, i.e., $(\mathbf{\hat{p}}_{1}^r, ..., \mathbf{\hat{p}}_{T}^r)|_{\epsilon} \sim \mathcal{Y}(\mathbf{p}^r_{1:T}, \epsilon)$. The generated synthetic forecasts $\mathbf{\hat{p}}_{t}^r$ aim at realistically mimicking the operational forecasts available to the controller at step $t$, e.g., delivered by the grid operator's prediction module, and are used as controller input for decision making during the control horizon.

Specifically, to generate synthetic forecast $\mathbf{\hat{p}}_{t}^r$, we propose adding synthetic forecast errors, which follow a carefully designed distribution, to the actual generation profile. The benefit for this approach is twofold: first, unlike other approaches, such as building a renewable generation forecast module based on time-series analysis, our approach can eliminate the impact of the forecasting module modeling error to the downstream analysis. Second, as discussed in Section \ref{subsec-state}, the renewable forecast accuracy has a strong impact on the CLR controller's performance, for both the optimization based and the proposed RL based. By adjusting the distribution of the added errors, we can mimic forecasts with different error levels, allowing us to quantitatively evaluate the robustness of these controllers, given different degrees of uncertainty.

Though we generate $\mathbf{\hat{p}}_{t}^r$ through synthesis, two principles are followed when designing $\mathcal{Y}$ to make $\mathbf{\hat{p}}_{t}^r$ realistic:

\noindent $\boldsymbol{[P.1]}$ Forecast error accumulates through time, making a $x$-step ahead forecast \textit{statistically less accurate than} a $x'$-step ahead forecast, given $x>x'$.

\noindent $\boldsymbol{[P.2]}$ Assume grid operator receives an updated forecast every control step. So, forecasts at two adjacent steps, i.e., $\mathbf{\hat{p}}_{t}^r$ and $\mathbf{\hat{p}}_{t+1}^r$, are highly correlated, and $\mathbf{\hat{p}}_{t}^r$ is updated using the latest renewable generation realization $p^r_{t+1}$ to obtain $\mathbf{\hat{p}}_{t+1}^r$.

\noindent With these two principles, the following two subsections describe the process of $\mathcal{Y}$.

\subsection{Generation of Synthetic Forecasts with Given Error Levels}

\begin{figure}[]
\centering
\includegraphics[width=0.95\linewidth]{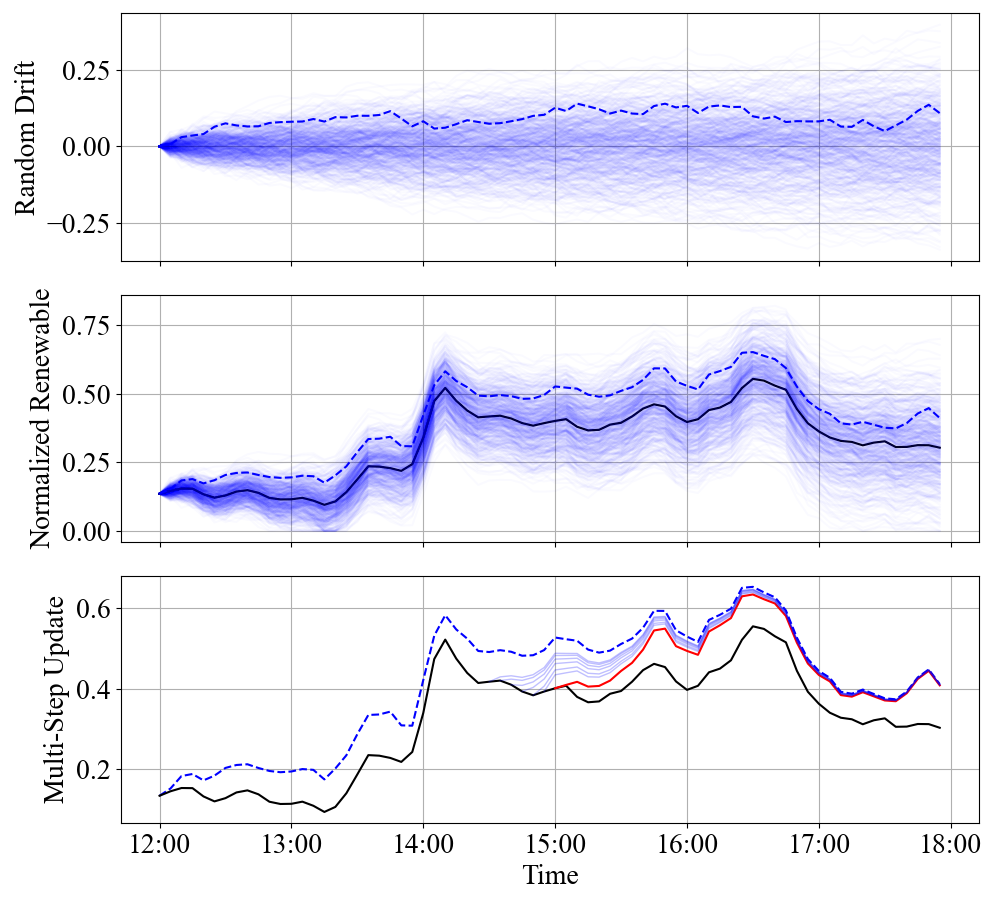}
\caption{Example of synthetic forecasts of wind generation with $\epsilon_T=0.1$. \textit{Top}: Sampled 500 random drifts over time generated based on Proposition \ref{proposition-1}, i.e., $\sum_{i=0}^t \iota_i$. \textit{Middle}: Corresponding synthetic forecasts (all normalized), i.e., $\Gamma_{[0,1]}(p_{k}^r /p^{r,\text{max}} + \sum_{i=1}^k \iota_i )$, given the 500 random drifts. Black curve is the actual generation $\mathbf{p}_{1:T}^r$, and the synthetic forecasts $\mathbf{\hat{p}}_{1:T}^r$ could be any one of the blue curves, the dashed line shows one specific over-forecast sample. \textit{Bottom}: Red curve shows $\mathbf{\hat{p}}_{t:T}^r$ at a point and the trailing faded blue curves are forecasts at previous steps, i.e., $\mathbf{\hat{p}}_{t-i:T}^r, i \in \{1, ..., 5\}$. It is similar for PV synthetic forecasts generation.}
\label{fig-pfg-example}
\end{figure}

In this study, we use the \textit{expected T-step prediction relative error}, i.e., $E_T^{\text{pred}}$, as the parameter to control the forecast error level. Concretely, for a forecast horizon $T$ and provided $\mathbf{p}^r_{1:T} = [p_1^r, p_{2}^r, ..., p_{T}^r]^\top$, the first step synthetic forecast $\mathbf{\hat{p}}_{1:T}^r = [p_1^r, \hat{p}_{2|1}^r, ..., \hat{p}_{T|1}^r]^\top$ is generated such that because of the accumulation of multi-step forecast errors, $E_T^{\text{pred}}=\mathbb{E}[|p_{T}^r - \hat{p}_{T|1}^r| / p^{r, \text{max}}]$ is equal to a pre-set level, e.g., $\epsilon_T=0.1$, indicating 10\% expected relative error. $p^{r, \text{max}}$ here is the generation capacity of renewable DER $r$.

\begin{proposition}  \label{proposition-1}
If the single step normalized forecast errors, $\iota_i$, $\forall i \in \mathcal{T}$, follow $\iota_i \sim N(0, (\epsilon_T \sqrt{\frac{\pi}{2T}})^2)$ i.i.d, and errors accumulate, then $E_T^{\text{pred}}=\epsilon_T$. Further, the end-of-horizon standard deviation of the expected relative error is $\epsilon_T \sqrt{\frac{\pi}{2} - 1}$.
\end{proposition}

\noindent\textit{Proof.} See Appendix \ref{appen-proof-proposition-1}. $\hfill \qed$

Following Proposition \ref{proposition-1}, $\mathbf{\hat{p}}_{1:T}^r := [p_1^r, \hat{p}_{2|1}^r, ..., \hat{p}_{T|1}^r]^\top$ is generated by
\begin{equation}  \label{eq-pseudo-forecast-generation}
    \hat{p}_{j|1}^r = \Gamma_{[0, c^r_j]}(p_{j}^r + \sum_{i=1}^j \iota_i p^{r, \text{max}}), \quad j \in \{2, 3,..., T\}
\end{equation}
where $c^r_j$ is the maximum possible renewable generation at step $j$. For PV system, $c^r_j \leq p^{r, \text{max}}$ is time-variant and is given by the clear sky generation at step $j$; and for wind generation forecast, there is $c^r_j = p^{r, \text{max}}$ constantly. The projection $\Gamma_{[0, c^r_j]}(y)=\underset{y' \in [0, c^r_j]}{\mathrm{argmin}} ||y'-y||$ ensures feasible forecasts. See Fig. \ref{fig-pfg-example} (\textit{Top} and \textit{Middle}) for an example.

\begin{figure}[]
\centering
\includegraphics[width=0.95\linewidth]{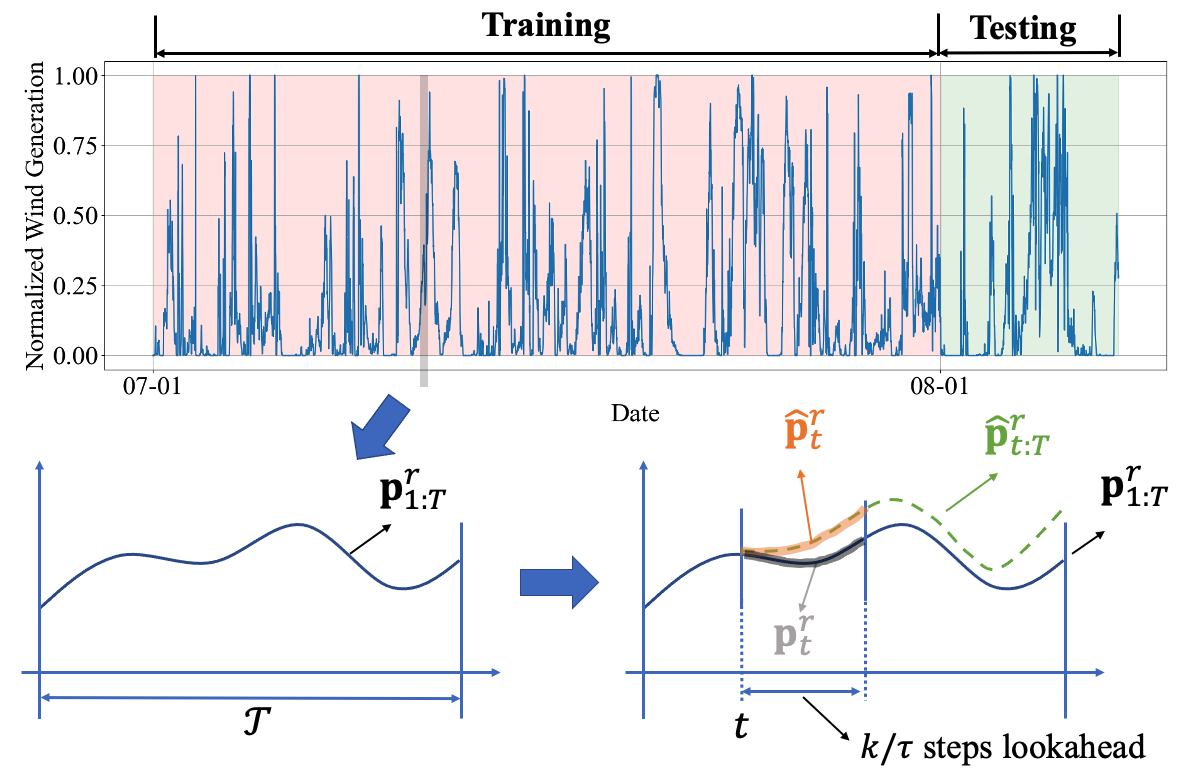}
\caption{Illustration of scenario sampling and synthetic forecasts generation. \textit{Top}: The renewable generation of one DER \cite{draxl2015wind}, denoted as $\mathcal{P}^r \subset \mathcal{P}^\mathcal{R}$. \textit{Lower left}: The actual renewable generation profile sampled for one episode, i.e., $\mathbf{p}^r_{1:T} \sim \mathcal{P}^r$; \textit{Lower right}: This subplot highlights the relationship among $\mathbf{p}^r_{1:T} \in \mathbb{R}^{T}$, $\mathbf{\hat{p}}_{t:T}^r \in \mathbb{R}^{T-t+1}$ (updated forecasts at step $t$), $\mathbf{\hat{p}}_{t}^r \in \mathbb{R}^{k/\tau}$ (updated $k$-hour lookahead forecasts to be used in RL state \eqref{eq-renewable-forecasts}) and $\mathbf{p}_{t}^r \in \mathbb{R}^{k/\tau}$ ($k$-hour lookahead perfect forecasts to be used in \eqref{eq-stage-I-state-definition} in Section \ref{subsec-cl-framework}).}
\label{fig-scenario-and-state}
\end{figure}

\subsection{Forecasts Update within Control Horizon}  \label{subsec-forecast-updates}
As \textbf{P.2} indicates, for the rest of the steps in the control horizon, synthetic forecasts $\mathbf{\hat{p}}_{t:T}^r$ are updated every step based on the latest renewable generation realization. Specifically, given $\mathbf{\hat{p}}_{t:T}^r = [p_t^r, \hat{p}_{t+1|t}^r, ..., \hat{p}_{T|t}^r]^\top$ for $t\in[1, T-1]$, at step $t+1$, upon the realization of $p_{t+1}^r$, the forecasts, i.e., $\mathbf{\hat{p}}_{t+1:T}^r = [p_{t+1}^r, \hat{p}_{t+2|t+1}^r, ..., \hat{p}_{T|t+1}^r]^\top$, are updated as follows: 

\begin{multline}
  \hat{p}_{t+x|t+1}^r = \\ \hat{p}_{t+x|t}^r + \beta^{x-1}[p_{t+1}^r - \hat{p}_{t+1|t}^r], \quad \forall x \in [1, T-t]
\end{multline}
in which $\beta \in [0, 1]$ controls how far the impact this single step realization reaches towards future forecasts. In this study, we use $\beta=0.9$ to keep the impact within two hours ahead. See the red curve and the trailing faded blue curves in Fig. \ref{fig-pfg-example} (\textit{Bottom}) for an example.

\subsection{Generate training and testing data set}

Based on the mechanism introduced above, the training data set is generated as follows:

\begin{equation} \label{eq-train-data-set}
\begin{aligned}
\Xi_{\epsilon}^{\text{tr}} =\bigcup_{\forall r\in \mathcal{R}}^{\text{sync}} \{ &(\mathbf{p}^r_{1:T}, \mathbf{\hat{p}}_{1}^r, ..., \mathbf{\hat{p}}_{T}^r) \  | \  \mathbf{p}^r_{1:T} \sim \mathcal{P}^{r, \text{tr}}, \\ &(\mathbf{\hat{p}}_{1}^r, ..., \mathbf{\hat{p}}_{T}^r)|_{\epsilon} \sim \mathcal{Y}(\mathbf{p}^r_{1:T}, \epsilon)\},
\end{aligned}
\end{equation}
where $\mathcal{P}^{r, \text{tr}}$ is the actual renewable generation profile reserved for training, and correspondingly, there is $\mathcal{P}^{r, \text{ts}}$ for testing, see Fig. \ref{fig-scenario-and-state} (Top). The symbol $\bigcup_{\forall r\in \mathcal{R}}^{\text{sync}}$ means the generation profiles for all DERs are sampled synchronously to preserve the temporal correlation. For $t > T-k/\tau+1$, there will be $|\mathbf{\hat{p}}_{t:T}^r| < k/\tau$, in order to keep $|\mathbf{\hat{p}}_{t}^r|=k/\tau$ (since $\mathbf{\hat{p}}_{t}^r$ is a part of the RL state, which needs to be of fixed length), additional dummy padding, e.g., $\mathbf{1}_{k/\tau-T+t-1}$, is appended to $\mathbf{\hat{p}}_{t:T}^r$ to form $\mathbf{\hat{p}}_{t}^r$. The testing data set $\Xi_{\epsilon}^{\text{ts}}$ is generated in the same approach, though the actual renewable generation profile is sampled from $\mathcal{P}^{r, \text{ts}}$, so that the scenarios are different from those in $\Xi_{\epsilon}^{\text{tr}}$.

Finally, it is worth re-emphasizing that the methodology introduced in this section is just used to facilitate the investigation in this paper; in real-life application, $\Xi^{\text{tr}}$ consists of historical generation profiles and historical forecasts generated by the prediction module used by the grid operator. 

\section{Curriculum Learning based RL Training}

In this section, we will discuss how an RL controller can be trained based on the previously defined MDP and using the generated training scenarios $\Xi^{\text{tr}}_{\epsilon}$.

\subsection{RL Policy Searching and the Challenge}  \label{subsec-dl-challenges}

Among many RL algorithms, we choose policy-based ones, which directly search in the policy space to achieve the maximization in (\ref{eq-optimal-policy-def}). In deep RL, the policy is instantiated using a deep neural network, i.e., $\pi_{\bm{\psi}}(\mathbf{a}_t|\mathbf{s}_t)$, where ${\bm{\psi}}$ represents the parameters (weights and biases) of the policy network. Following a given policy $\pi_{\bm{\psi}}(\mathbf{a}_t|\mathbf{s}_t)$, the RL controller's performance can be written as $J(\bm{\psi})=\mathbb{E}_{\mathbf{a}_t\sim\pi_{\bm{\psi}}, \xi \sim \Xi^{\text{tr}}_{\epsilon}}(\sum_{t\in \mathcal{T}} r_t)$. Therefore, (\ref{eq-optimal-policy-def}) becomes searching optimal policy parameters, i.e., $\bm{\psi}^*=\argmax_{\bm{\psi}} J(\bm{\psi})$. To this end, a group of policy-based RL algorithms, use gradient ascent for policy update:
\begin{equation}  \label{eq-policy-gradient-ascent}
    \bm{\psi}_{k+1} = \bm{\psi}_{k} + \kappa \widehat{\nabla}_{\bm{\psi}} J(\bm{\psi})
\end{equation}
where $\widehat{\nabla}_{\bm{\psi}} J(\bm{\psi})$ is the policy gradient estimated from collected experience, $k$ is the learning iteration and $\kappa$ is the learning rate. Algorithms may estimate $\widehat{\nabla}_{\bm{\psi}} J(\bm{\psi})$ either based on \textit{policy gradient theorem} \cite[Chapter 13]{sutton2018reinforcement} (algorithms such as TRPO \cite{schulman2015trust} and PPO \cite{schulman2017proximal}) or other methods such as zero-order gradient estimation (algorithms like ES-RL \cite{salimans2017evolution}). However, as will be substantiated in Section \ref{sec-case-study}, when directly applying these algorithms to solve the problem depicted by (\ref{eq-opt-formulation}), the trained controllers fail to show desirable performance. This issue is originated from 1) the grid control problem complexity, 2) large policy search space and 3) the non-convexity of $J(\bm{\psi})$; all these combined makes it hard to set suitable hyper-parameters to properly explore the policy space and escape local optima in the non-convex policy search process. So, when randomly initialize the policy network $\pi_{\bm{\psi}}(\mathbf{a}_t|\mathbf{s}_t)$ of an RL agent and train it in a non-convex environment using a gradient method, converging to a sub-optimal local optimum is very likely.  

\subsection{Curriculum Learning (CL) Framework for CLR} \label{subsec-cl-framework}

We investigate leveraging CL to ameliorate such challenge by learning in a simpler steppingstone problem first to provide better initial policy for training to solve \eqref{eq-opt-formulation}, and ultimately facilitates a better policy convergence. Though deep learning researchers are investigating how to choose curriculum automatically \cite{portelas2020automatic}, we defer this to our future work and in this paper, the learning curriculum is designed based on domain knowledge. Namely, different aspects of complexity in the problem of interest that an RL agent needs to learn are first identified, and then select a subset of them to form the simplified problem. In the formulated CLR problem, the following two aspects represent the learning challenges:

\textbf{[C.1]} Complex relationships in grid control problems, e.g., generation-load balance in the islanded grid and system power flow (i.e., bus voltages vs. DERs set points relationship).

\textbf{[C.2]} Extract helpful information from renewable forecasts with error to maximize the load restoration.

Corresponding to these complexities, the following simplifications are made to form a simpler problem:

1) Reducing the action space: instead of learning to control both load pick-up and DERs set points, i.e.,  $\mathbf{p}_t$, $\mathbf{H}_p\mathbf{p}^\mathcal{G}_t$, and $\mathbf{H}_{\alpha}\bm{\alpha}^\mathcal{G}_t$, RL agent only determines the DERs dispatch:

\begin{equation} \label{eq-stage-I-action-definition}
    \mathbf{a}_t^I := [(\mathbf{p}^\mathcal{G}_t)^\top, (\bm{\alpha}^\mathcal{G}_t)^\top]^\top \in \mathcal{A}^I.
\end{equation}
Once the total generation is determined, loads are restored greedily according to their importance ($\zeta^i$), i.e., $\mathbf{p}_t^* = \argmax_{\mathbf{p}_t} \mathbf{z}^\top \mathbf{p}_t$, subject to $\mathbf{1}_N^{\top} \mathbf{p}_t = \mathbf{1}_{|\mathcal{G}|}^{\top} \mathbf{p}_t^{\mathcal{G}}$ and $\mathbf{0}_N \leq \mathbf{p}_t \leq \mathbf{p}$. Restoring loads in this greedy way leads to solutions that are feasible to the original problem but might be sub-optimal. This is because if restoring a load causes voltage violation, then all other loads with lower $\zeta^i$ will not be restored, even if restoring them will not cause any voltage violation. This limits the amount of load to be restored, and thus only gives sub-optimal solutions.

2) At each control step $t$, informing the RL agent with errorless/perfect forecasts $\mathbf{p}_{t}^{\mathcal{R}}=[(\mathbf{p}_{t}^1)^\top, ..., (\mathbf{p}_{t}^{|\mathcal{R}|})^\top]^\top$, i.e., actual renewable generation in the lookahead period, instead of $\hat{\mathbf{p}}_{t}^{\mathcal{R}}$ (See Fig. \ref{fig-scenario-and-state} for an illustration regarding the difference between $\mathbf{p}_{t}^r$ and $\hat{\mathbf{p}}_{t}^r (r \in \mathcal{R})$). This leads to a slightly modified RL state as follows:

\begin{equation}  \label{eq-stage-I-state-definition}
\mathbf{s}_t^I := [(\mathbf{p}_{t}^{\mathcal{R}})^\top, (\widetilde{\mathbf{p}}_{t-1})^\top, (\mathbf{S}_t^{\theta})^\top, (\mathring{\mathbf{E}}_t^{\mu})^\top, t, (\Phi_t)^\top]^\top,
\end{equation}
allowing the RL agent to focus on learning grid control problem without uncertainty in this simpler version.

As a result, a two-stage CL framework is devised: in Stage I, the RL agent learns to solve the simplified problem and once a good policy is trained, the knowledge is transferred to Stage II to jump-start the policy training for the original problem. 

\begin{figure}[]
\centering
\includegraphics[width=0.98\linewidth]{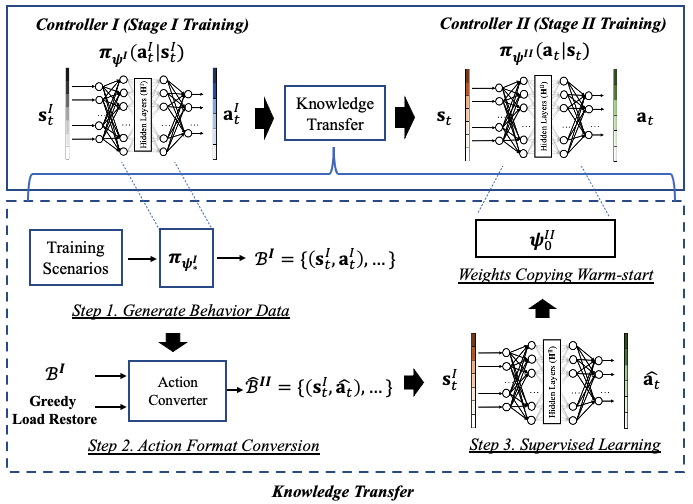}
\caption{Curriculum learning framework overview.}
\label{fig-knowledge-transfer}
\end{figure}

\subsection{Knowledge Transfer Between Stages}
One remaining difficulty is regarding the knowledge transfer between Stage I and II, because policy networks in two stages have different structure ($\mathcal{A}\neq \mathcal{A^I}$), which makes direct weights copying invalid. To tackle this, we propose a behavior cloning technique which trains a policy network with Stage II format but clones the Stage I trained controller's behavior on $\Xi^{\text{tr}}$, and then use this network to initiate the Stage II training. See Algorithm \ref{algo-curriculum-learning} for details regarding transferring knowledge between two heterogeneous policy networks. The overall two-stage CL workflow is illustrated in Fig. \ref{fig-knowledge-transfer}.

\begin{algorithm}
\caption{Curriculum-based RL for the CLR Problem}
\label{algo-curriculum-learning}
\begin{algorithmic}[1]
\renewcommand{\algorithmicrequire}{\textbf{Input:}}
\REQUIRE {All parameters for the CLR problem (\ref{eq-opt-formulation}) and randomly initialized Stage I policy parameters $\bm{\psi}_0^I$}

\STATE \textbf{Stage I Learning} \\

Train an RL control policy $\pi_{\bm{\psi}_*^I}(\mathbf{a}^I_t|\mathbf{s}^I_t)$ to solve the simplified CLR problem.

\STATE \textbf{Knowledge Transfer} \\

\textbf{\textit{Step 1}}: Apply $\pi_{\bm{\psi}_*^I}(\mathbf{a}^I_t|\mathbf{s}^I_t)$ in training scenarios $\Xi^{\text{tr}}$ to generate state-action pair from control trajectories $ \mathcal{B}^I := \{(\mathbf{s}_t^I, \mathbf{a}_t^I), ...\}$. \\

\textbf{\textit{Step 2}}: Convert the format of $\mathbf{a}_t^{I}$ in $\mathcal{B}^I$ to that of $\mathbf{a}_t$, and obtaining $\widehat{\mathcal{B}}^{II} := \{(\mathbf{s}_t^I, \widehat{\mathbf{a}_t}), ...\}$. The load pickup component $\mathbf{p}_t$ in $\widehat{\mathbf{a}_t}$ reflects the greedy restoration behavior. \\

\textbf{\textit{Step 3}}: Train a neural network that maps $\mathbf{s}_t^I$ to $\widehat{\mathbf{a}_t}$ via supervised learning using $\widehat{\mathcal{B}}^{II}$ as a training data set, obtaining trained network parameters as $\bm{\psi}_0^{II}$. \\

\STATE \textbf{Stage II Learning} \\

Initialize an RL control policy for the original problem with parameters $\bm{\psi}_0^{II}$, i.e., warm-start the policy as $\pi_{\bm{\psi}_0^{II}}(\mathbf{a}_t|\mathbf{s}_t)$, and continue to train it until converged as $\pi_{\bm{\psi}_*^{II}}(\mathbf{a}_t|\mathbf{s}_t)$.

\RETURN {Two-stage curriculum trained policy $\pi_{\bm{\psi}_*^{II}}(\mathbf{a}_t|\mathbf{s}_t)$} 
\end{algorithmic}
\end{algorithm}

\section{Case Study} \label{sec-case-study}

\subsection{Experiment Setup} \label{subsec-case-study-setup} 

The proposed CL framework is investigated using modified IEEE 13-bus and IEEE 123-Bus system. 

\subsubsection{IEEE 13-Bus Test System}
\begin{figure}[]
\centering
\includegraphics[width=0.9\linewidth]{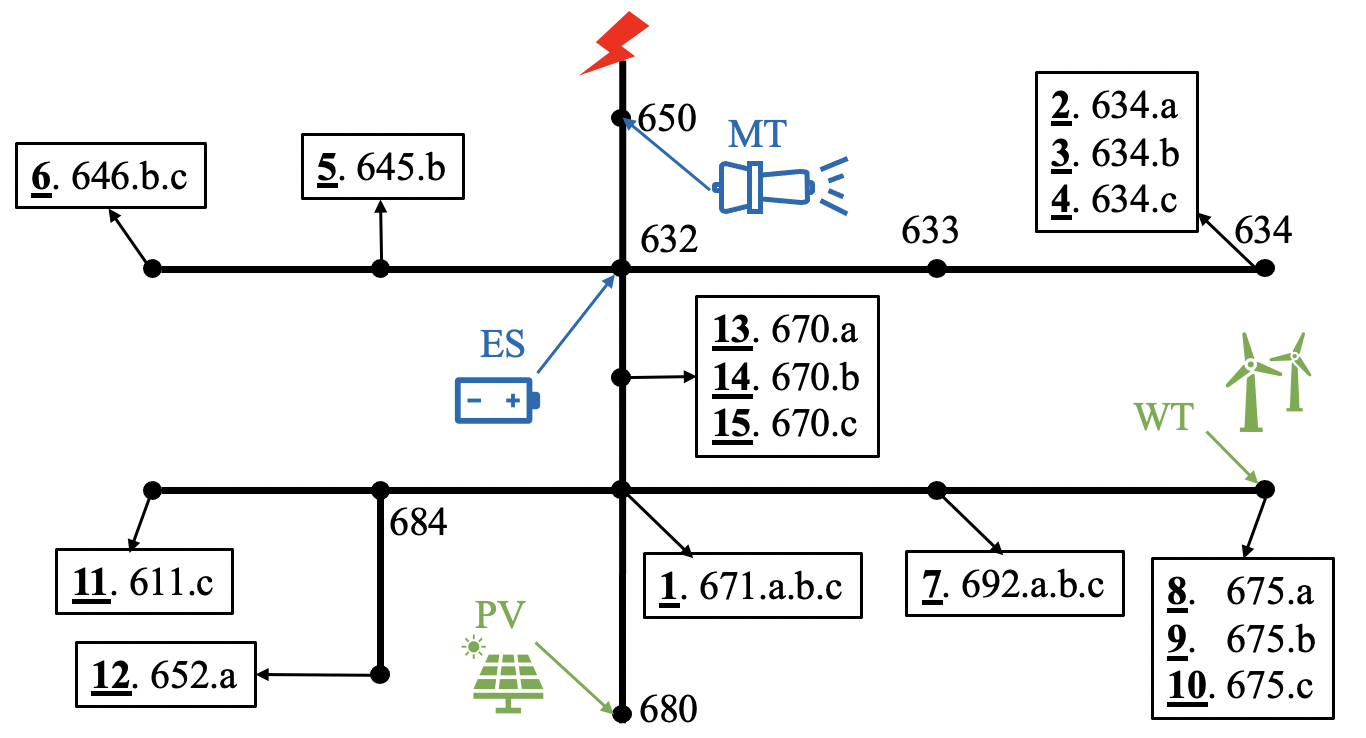}
\caption{A testing system modified from the IEEE 13-bus system.}
\label{fig-13-bus-illustration}
\end{figure}

As illustrated in Fig. \ref{fig-13-bus-illustration}, four DERs ($|\mathcal{G}|=4$) and a total of 15 single-/multi-phase critical loads ($|\mathcal{L}|=15$) are considered, with their parameters summarized in Table \ref{table-DER-configuration} ($N_{trunc}$ in Table \ref{table-DER-configuration} indicates the truncated normal distribution, and the units for power and energy are respectively kW and kWh). In this study, we assume the load restoration duration is known, and a six-hour horizon is used with five minute control intervals, i.e., $\tau = 1/12$ and $\mathcal{T}=\{1,2,...,72\}$. Load priority factor $\epsilon^i, i\in\mathcal{L}$ is set to be in $[0.2, 1.0]$ to indicate relative importance comparison among loads. Voltage limits are set to be $V^{\text{max}}=1.05$ p.u. and $V^{\text{min}}=0.95$ p.u., and the voltage violation unit penalty is set to $\lambda=10^8$.

\begin{table}[]
\centering
\caption{Parameters of DERs and Critical Loads}
\label{table-DER-configuration}

\begin{tabular}{cc}
\specialrule{.15em}{.075em}{.075em}

\textbf{Entity} & \textbf{Parameters} \\ \hline
\multirow{2}{*}{\textit{Micro-Turbine} ($\mu$)}     & $p^{\mu}\in[0,400]$, $E^{\mu}=1200$ \\
& $\alpha^{\mu} \in [0, \pi/4]$ \\ \hline
\multirow{3}{*}{\textit{Energy Storage} ($\theta$)} & $-p^{\theta, ch}=p^{\theta, dis}=250$ \\ 
& $160 \leq S_{t}^{\theta} \leq 1250$, $\alpha^{\theta} \in [0, \pi/4]$ \\ & $s_0\sim N_{trunc}(1000, 250^2, 750, 1250)$ \\ \hline
\textit{PV} ($\rho$)                       & $p^{\rho} \in [0, 300]$, $\alpha^{\rho} \in [0, \pi/4]$ \\ \hline
\textit{Wind} ($\omega$)                       & $p^{\rho} \in [0, 400]$, $\alpha^{\omega} \in [0, \pi/4]$ \\ \hline
\multirow{3}{*}{\textit{Load} ($\mathcal{L}$)}  & $\epsilon^i=100, \forall i\in \mathcal{L}$ \\ & $\mathbf{z} = [1.0, 1.0, 0.9, 0.85, 0.8, 0.8, 0.75, 0.7$,\\ & $0.65, 0.5, 0.45, 0.4, 0.3, 0.3, 0.2]^\top$ \\
\specialrule{.15em}{.075em}{.075em}

\end{tabular}
\end{table}

\subsubsection{Learning Environment}  \label{subsubsec-learning-env}

\begin{figure}[]
\centering
\includegraphics[width=1.0\linewidth]{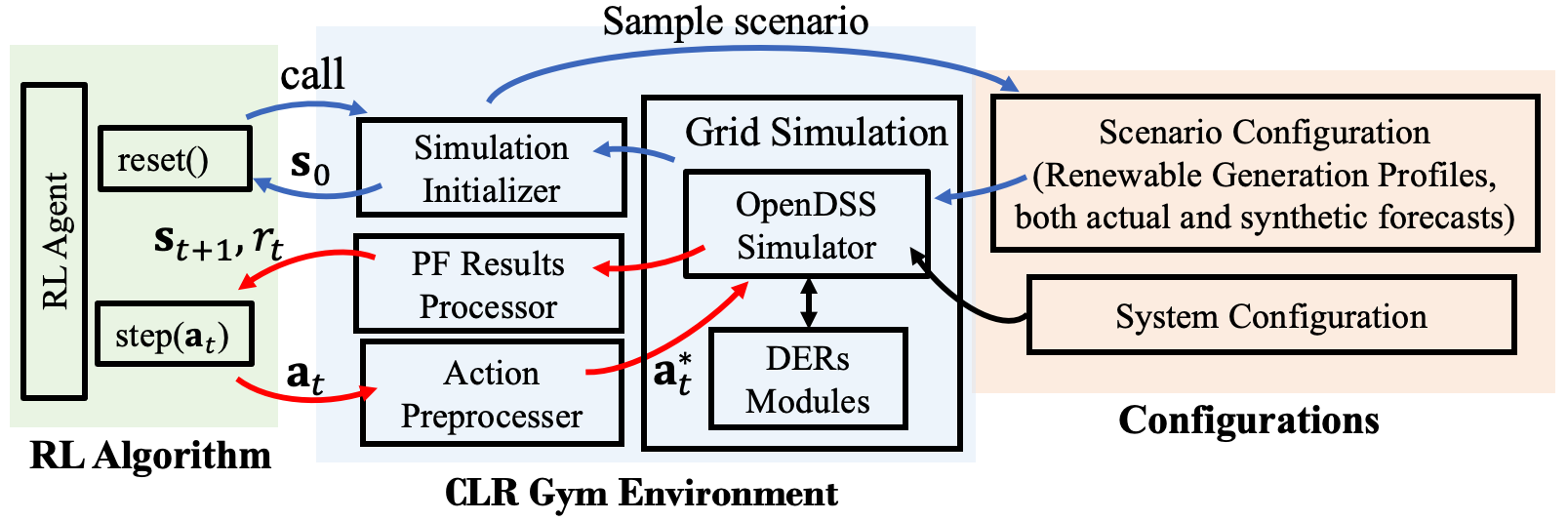}
\caption{Learning environment for the CLR problem. Using the ``reset'' and ``step'' interfaces, the RL agent can sample different $T$-step control scenarios and explore and learn the optimal control strategy.}
\label{fig-gym-structure}
\end{figure}

Following a standard OpenAI Gym interface \cite{brockman2016openai}, a learning environment, illustrated in Fig. \ref{fig-gym-structure}, is developed. The role of this environment is twofold: first, it provides programming interface between the RL agent and the grid simulator, i.e., OpenDSS; second, it is also responsible for enforcing some operational constraints, by projecting the agent's unconstrained action to a feasible one. Recall the penalty term method for handling the voltage constraint, letting the environment to enforce constraints is another approach we use to guarantee the action feasibility.
Specifically, below are several types of constraints that are enforced this way: 

1) \textit{Box constraints for individual element in $\mathbf{a}_t$}, e.g., $p_t^i \in [0, p^i]$ and $p_t^g\in[0, \overline{p^g}]$: For such constraints, outputs from the policy network are first clipped to $[-1, 1]$ and then mapped to variables' corresponding feasible range, e.g., $[0, p^i]$ or $[0, \overline{p^g}]$.

2) \textit{Power balance constraints}, i.e., $\mathbf{1}_N^{\top} \mathbf{p}_t = \mathbf{1}_{|\mathcal{G}|}^{\top} \mathbf{p}_t^{\mathcal{G}}$: This constraint places an equality relationship among multiple outputs of the policy network (e.g., $\mathbf{p}_t$ and $\mathbf{p}_t^{\mathcal{G}}$ in $\mathbf{a}_t$). To enforce this, our approach is to map $\mathbf{a}_t$ to a feasible $\mathbf{a}^*_t$ using a rule-based mechanism: namely, decrease $\mathbf{p}_t^{\mathcal{G}}$ if $\mathbf{1}_N^{\top} \mathbf{p}_t < \mathbf{1}_{|\mathcal{G}|}^{\top} \mathbf{p}_t^{\mathcal{G}}$ and reduce $\mathbf{p}_t$ otherwise. For details, see \cite[Algorithm 1]{zhang2021restoring}, which is not included here in the interest of space.

3) \textit{Resource availability constraints}, e.g., $\sum_{t \in \mathcal{T}} p_t^{g} \cdot \tau \leq E^{g}$. If the resource is depleted at step $k$, i.e., $\sum_{t =1}^{k} p_t^{g} = E^{g}$, the environment will forces $p_{t}^g=0$ for $t>k$ no matter what the corresponding value in $\mathbf{a}_t$ is.

The ``Action Preprocessor'' in Fig. \ref{fig-gym-structure} shows the above-mentioned projection in the environment implementation that ensures the action feasibility. In addition to these constraints, some equality constraints, e.g., battery SOC dynamics and power flow constraint, are naturally handled by the learning environment: for example, the environment directly calculates $S_{t+1}^{\theta}$ using the given SOC dynamics $S_{t+1}^{\theta} = S_{t}^{\theta} - \eta_t \cdot p_t^{\theta} \cdot \tau$, so such constraints are directly satisfied.

The ``Scenario Configuration'' block in Fig. \ref{fig-gym-structure} collects scenarios $\Xi^{\text{tr}}$ or $\Xi^{\text{ts}}$. During training, the learning environment repeatedly samples $\xi \sim \Xi^{\text{tr}}$ and lets the RL agent learn $\pi^*(\mathbf{a}_t|\mathbf{s}_t)$ via \eqref{eq-optimal-policy-def}. Once adequately trained, the RL agent has been exposed to various outage scenarios and thus can conduct proper control. In this study, 30 days of actual renewable generation profiles are used to generate $\Xi^{\text{tr}}$ for training, and data for the following seven days are used for generating $\Xi^{\text{ts}}$ (as illustrated in Fig. \ref{fig-scenario-and-state} (\textit{Top})). 
In real life, due to the difference in locations, the nature of resources and the prediction module used by grid operators, renewable forecasts at different generation site might have different error levels. To examine the RL controller's behaviors under different error levels: we consider $\epsilon_T \in \mathcal{E}=\{0, 5\%, 10\%, 15\%, 20\%, 25\%\}$. Note that for each error level $\epsilon_T$, its training scenario $\Xi^{\text{tr}}_{\epsilon_T}$ is generated and one RL agent is trained before testing using $\Xi^{\text{ts}}_{\epsilon_T}$. Finally, to compare how different look-ahead length for renewable forecasts will impact the control behavior, we investigate $k\in\{1,2,4,6\}$, where $k$ is the number of hours per~\eqref{eq-renewable-forecasts}. In the following sections, RL-$k$ indicates the RL agent trained with renewable forecasts of length $k$.

\subsection{RL Training}

To train RL control policies for the two-stage learning in the proposed curriculum, we take advantage of two different RL algorithms: for the first stage, an evolution strategies based direct policy search method (ES-RL) \cite{salimans2017evolution} is used due to its scalability; and PPO \cite{schulman2017proximal} is leveraged in the Stage II training since it has a better local policy search performance. A similar practice is discussed in \cite{zhang2020grid}. Both algorithms are on-policy and following \eqref{eq-policy-gradient-ascent} to iteratively update the policy. ES-RL approximates $\widehat{\nabla}_{\bm{\psi}} J(\bm{\psi})$ using zero-order estimation while PPO does it using the policy gradient theorem (calculated using samples collected in current training batch). The software implementation we used is based on \cite{liang2018rllib}. In this study, RL policy training is conducted on the high performance computing (HPC) system located at the U.S. National Renewable Energy Laboratory. Each HPC computing node is equipped with a 36-core CPU and multi-node training is used when beneficial, e.g., Stage I. See Section \ref{subsec-larger-system} for details on computational requirements. The policy network used has hidden layers structure as [256, 256, 128, 128, 64, 64, 38] and \textit{tanh} as activation function. In addition, for numerical reasons during training, the reward is scaled by 0.001. 

\begin{figure} []
    \centering
  \subfloat[Stage I learning curves. \label{fig-stage-1-learning-curves}]{%
       \includegraphics[width=0.99\linewidth]{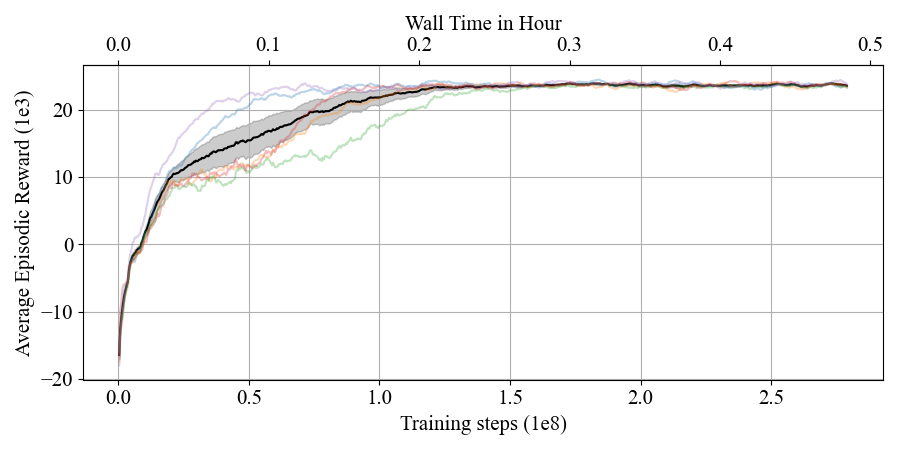}}
    \\
  \subfloat[Stage II learning curves for different $\epsilon_T$. \label{fig-stage-2-learning-curves}]{%
        \includegraphics[width=0.99\linewidth]{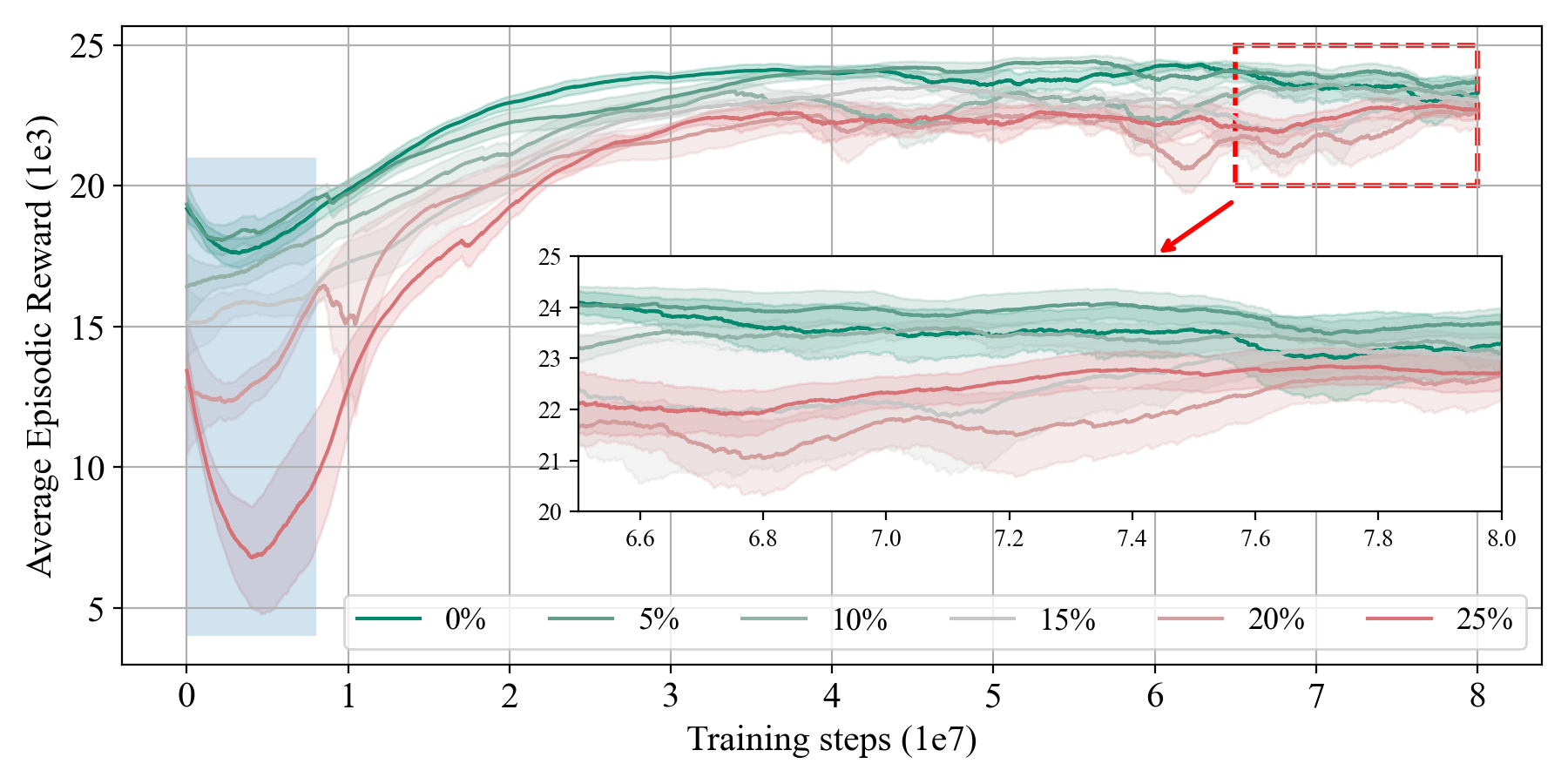}}

\caption{Learning curves for $k=1$ in Stage I (a) and Stage II (b) of the learning curriculum, demonstrating the controller performance improves as training progresses. In (a), the black curve and shaded area are the mean and standard deviation of reward from multiple separate experiments (faded color curves). Similarly, curves in (b) are averaged from multiple experiments and the blue shaded box shows the \textit{value function correction} phase.}
\label{fig-rl-learning-curves} 
\end{figure}

The learning curves of both stages are shown in Fig. \ref{fig-rl-learning-curves}. Fig. \ref{fig-stage-1-learning-curves} shows the Stage I learning for the simpler CLR problem over separate experiments: despite slight difference in converging process, the overall increase in reward level indicates effective learning. 
After transferring knowledge as described in Section \ref{subsec-cl-framework}, in Stage II, the $\mathbf{a}_t$, instead of $\mathbf{a}^I_t$, is used and renewable forecast error is introduced, considering all six cases with $\epsilon_T\in\mathcal{E}$. According to Fig. \ref{fig-stage-2-learning-curves}, two observations for Stage II training are made:

1) Although the converged reward at Stage I is around 23.5, applying the Stage I learned policies in Stage II environment causes lower reward (see $t=0$ values in Fig. \ref{fig-stage-2-learning-curves}) since the forecasts now is no longer perfect, and the larger $\epsilon_T$ is, the worse the starting performance.

2) Many curves show a decrease in reward at the early phase, as highlighted by the blue shaded box. This is because during the knowledge transfer, though the control behavior (i.e., actor network) is copied, the last layer of the PPO critic network is randomly initialized, which fails to provide accurate value estimation. Using these inaccurate estimations to update the actor, inevitably, deteriorates the control policy. However, once the critic is corrected, reward level starts to go up again and eventually converges to a higher reward level.

Next, we test these CL trained RL controllers.

\subsection{Comparison of RL and MPC under Uncertainty}  \label{subsec-rl-mpc-compare}

\begin{figure}[]
\centering
\includegraphics[width=1.0\linewidth]{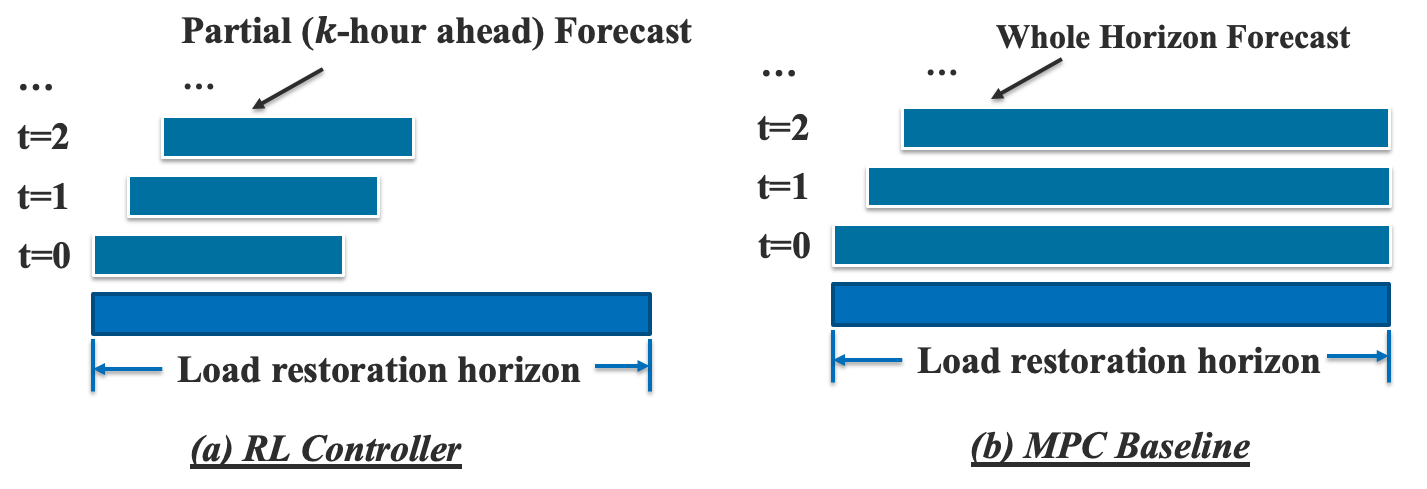}
\caption{Illustration of how RL and MPC controllers use renewable forecasts. Forecasts are updated every control interval as explained in Section \ref{subsec-forecast-updates}.}
\label{fig-controllers-comparison-illustration}
\end{figure}

\begin{figure}[]
\centering
\includegraphics[width=1.0\linewidth]{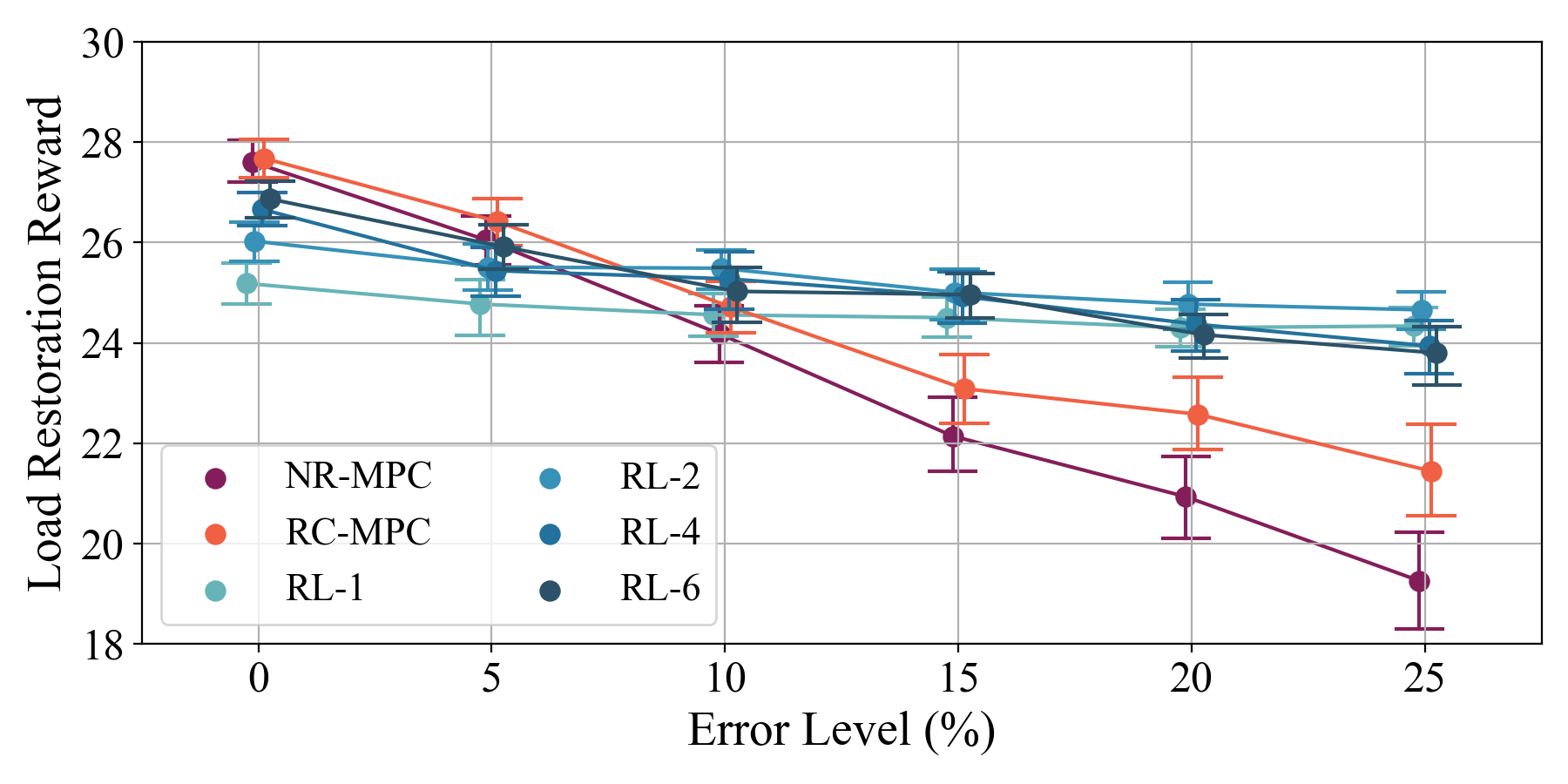}
\caption{Comparison between RL controller and MPC controllers under different $\epsilon_T$ regarding the load restoration reward, i.e., $\sum_{t\in \mathcal{T}} r_t^{\text{CLR}}$. Dots and vertical bands show the mean rewards over all testing scenarios and their corresponding 95\% confidence intervals. It shows RL controllers perform better in cases where the renewable forecasts are less accurate.}
\label{fig-rl-mpc-comparison}
\end{figure}

\begin{figure*} 
    \centering
    \subfloat[\textbf{Case I}. Low Forecast Error with $\epsilon_T = 5\%$\label{1a}]{%
       \includegraphics[width=0.325\linewidth]{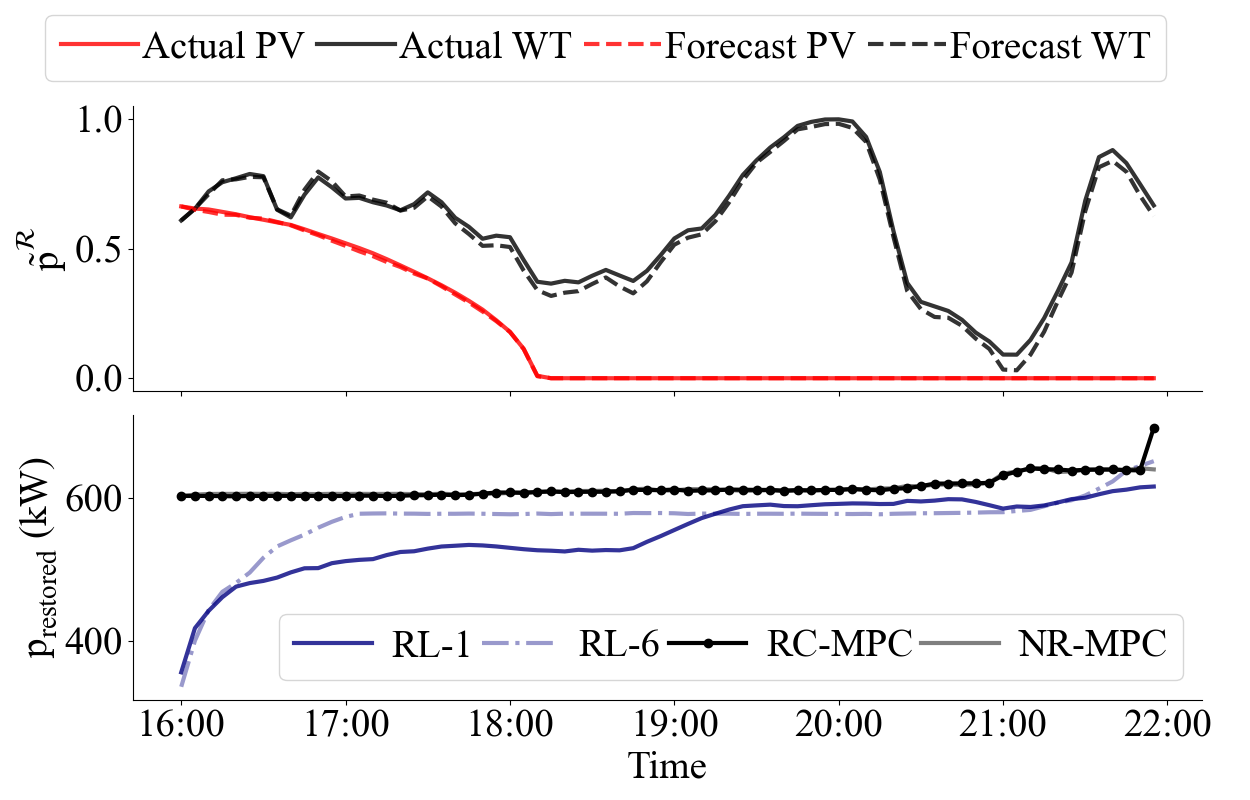}}
    \subfloat[\textbf{Case II}. Under-forecasting with $\epsilon_T = 25\%$\label{1b}]{%
        \includegraphics[width=0.325\linewidth]{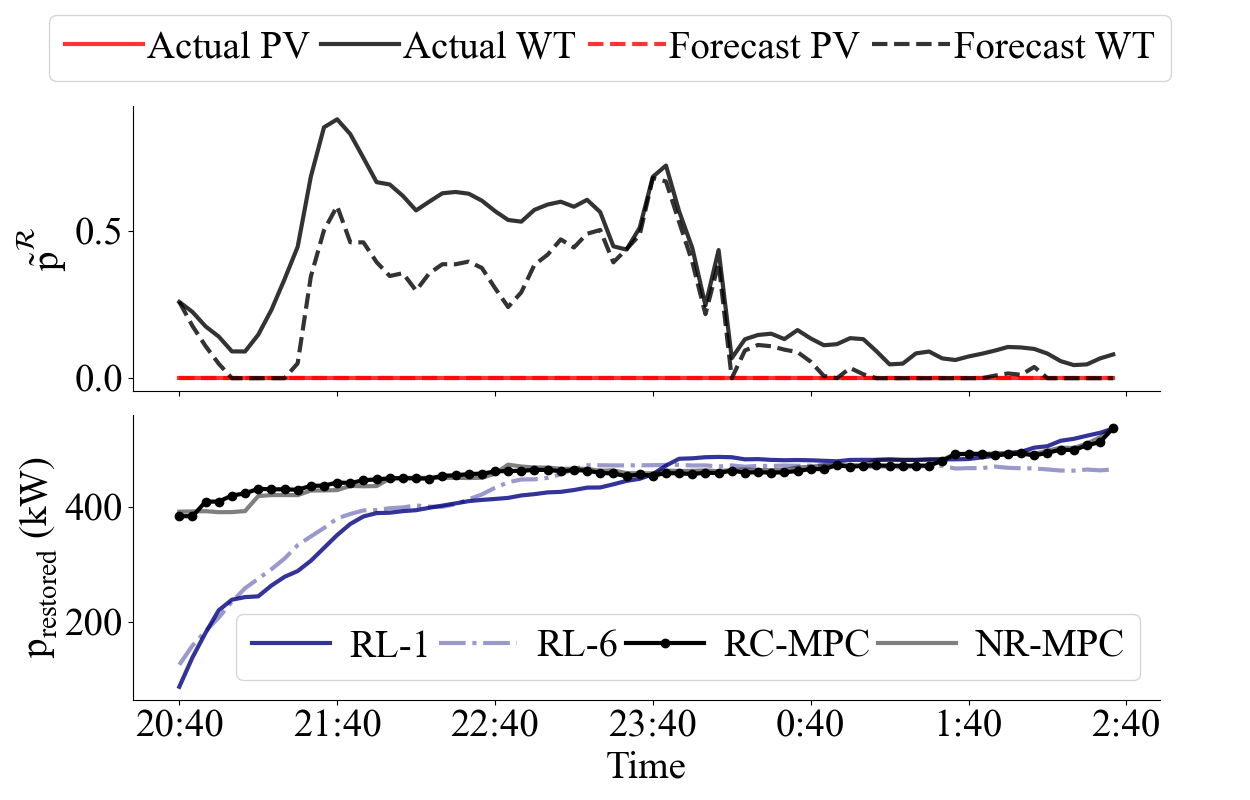}}
    \subfloat[\textbf{Case III}. Over-forecasting with $\epsilon_T = 25\%$\label{1c}]{%
        \includegraphics[width=0.325\linewidth]{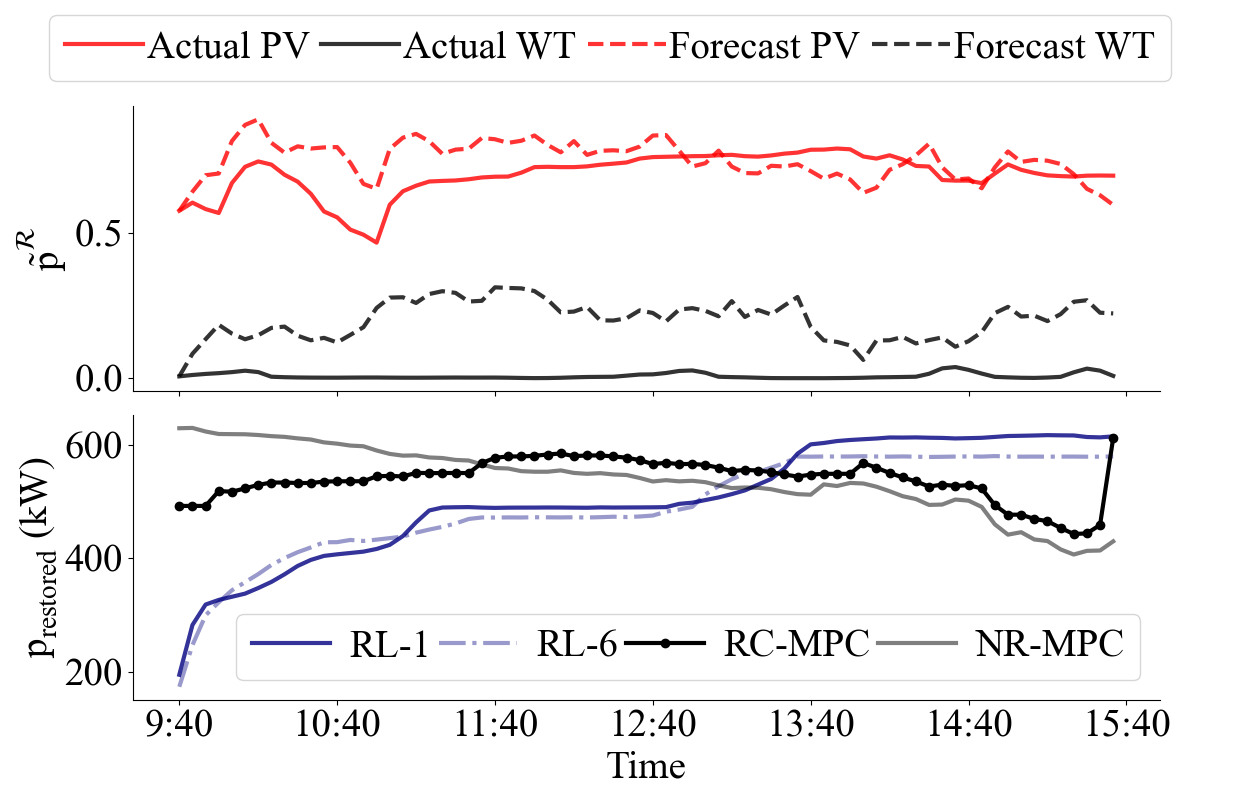}}
    \caption{Comparison of control behavior of RL controllers (showed two extremes: RL-1 and RL-6) and MPC baselines, under three testing scenarios. In three subplots, top plots show actual generation profiles $\mathbf{p}^r$ and their initial forecasts $\mathbf{\hat{p}}_1^r$; bottom plots show system restored power $\mathbf{1}_N \mathbf{p}_t$ over $\mathcal{T}$.}
  \label{fig-three-scenarios-comparison}
\end{figure*}

In this section, the trained RL controllers are compared with two MPC baselines using different $\Xi_{\epsilon_T}^{\text{ts}}$. 
Fig. \ref{fig-controllers-comparison-illustration} shows the inputs these two types of controllers used for decision making. 
It's worth noting that, for a fair comparison, $\Xi_{\epsilon_T}^{\text{ts}}$ has not been seen by the RL controller during training; as a result, the experiment here demonstrates how well the RL controller can be generalized. Total testing scenarios number is $504=3 \times 24 \times 7$ for each $\epsilon_T$ value, as we choose three scenarios per hour for seven testing days. 

\subsubsection{Load Restoration Performance}  \label{subsubsec-load-restoration}

Fig. \ref{fig-rl-mpc-comparison} shows the load restoration reward $\sum_{t\in \mathcal{T}} r_t^{\text{CLR}}$ distributions of different controllers under $\Xi_{\epsilon_T}^{\text{ts}}$, $\forall \epsilon_T \in\mathcal{E}$, and there are following observations and interpretations:

a) As expected, all controllers perform better if $\epsilon_T$ is small.

b) With $\epsilon_T=0$, model-based MPC controllers plan with perfect forecasts and thus the reward achieved provides a performance upper bound (e.g., 27.609 for NR-MPC). Using this as a baseline, it shows that RL controllers, even though model-free and without the optimality guarantee in policy search, can achieve a very good performance: 25.186 for RL-1 and 26.869 for RL-6, which represents 91.2\%-97.3\% of the performance upper bound. 

c) When $\epsilon_T$ is small (e.g., 0 or 5\%), $\{\mathbf{\hat{p}}_{t}^r, t\in\mathcal{T}\}$ are quite reliable, MPC baselines outperform RL controllers with its model-based nature and the optimality guarantee by the optimization-based method. But in cases with larger $\epsilon_T$, RL controllers are less susceptible to forecasting error than MPC, yielding higher reward. This is because RL controllers have learned from training experience that $\{\mathbf{\hat{p}}_{t}^r, t\in\mathcal{T}\}$ are unreliable, and thus develop a strategy to maximally leverage the unreliable forecasts.

d) Among RL controllers, RL-1 and RL-2 achieve worse performance than RL-4 and RL-6 when $\epsilon_T$ is small as they disregard reliable forecasts beyond one or two hours. On the other hand, RL-4 and RL-6 perform worse than RL-1 and RL-2 when $\epsilon_T$ becomes larger because the extra forecast information does not provide as much benefit due to increase error and the increased state space makes the controller harder to train.

To examine controllers' behavior in more details, Fig. \ref{fig-three-scenarios-comparison} compares the load restoration performance in three representative testing cases:

a) \textbf{Case I}: When $\epsilon_T$ is small, MPC uses the almost perfect forecasts and generates the restoration plan that is reliable and near optimal. In contrast, because of the lack of optimality guarantee in the RL training, the restoration performance by RL controllers is inferior though still reliable (i.e., monotonical load restoration). Specifically, RL-1, using only the immediate one-hour forecasts and discarding all future forecasts, acts even more cautiously and sub-optimally than RL-6.

b) \textbf{Case II \& III}: When $\epsilon_T$ is large, especially when the actual renewable generation is below the predicted, MPC needs to shed previously restored critical loads as it realizes the generation shortage during the process. This is also true in the RC-MPC case, in which a fixed reserve is considered. In contrast, RL controllers learn through training experience to distrust the provided forecasts and form restoration policies that are more robust to the forecast error.

\subsubsection{Voltage Constraining Performance}  \label{subsubsec-voltage-constraints}

For one episode, i.e., $t\in\mathcal{T}$, the following two performance metrics are defined to quantify violation duration and extent:

\begin{flalign}
    &\tau_{\text{vio}} := \sum_{i \in N_b, t\in\mathcal{T}} \mathds{1}_{\mathcal{V}_{\text{vio}}}(v_t^i) \tau,  \quad  \Tilde{V}_{\text{vio}} := \frac{1}{|\mathcal{V}_{\text{vio}}|} \sum_{v \in \mathcal{V}_{\text{vio}}} v&
\end{flalign}
in which $\mathcal{V}_{\text{vio}}:=\{v_t^i | v_t^i \notin [V^{\text{min}}, V^{\text{max}}], \forall t \in \mathcal{T}, \forall i \in N_b\}$ and the indicator function $\mathds{1}_{\mathcal{X}} (x)$ equals to 1.0 if $x\in \mathcal{X}$, otherwise zero. Table \ref{table-voltage-violation} shows the metrics evaluation under different $\epsilon_T$, $\overline{(\cdot)}$ is average over $\Xi^{\text{ts}}$ in which there is voltage violation. Because MPC controllers use a linearized power flow model in the optimization formulation, but the control is implemented on OpenDSS; in contrast, the RL controller can directly learn from the more accurate nonlinear power flow model, the linearization error in MPC formulation leads to more voltage violation (lower violated voltage) and longer duration/occurrence than RL controllers have. In other words, RL controllers suffer less from the modeling error, especially the linearization error, which is common in many optimization-based approach; though they still suffer if the OpenDSS model itself deviates from the actual grid being controlled.

Note, though voltage violation is rare and minor when using RL controller, the RL controller considers it as a soft constraint, i.e., (\ref{eq-obj-func}) and (\ref{eq-vv-penalty}). This does not guarantee no voltage violation, because if the voltage violation is small, the $\vartheta_t$ penalty in (\ref{eq-obj-func}) is insufficient to incentivize a policy update to the direction in which voltage violation is eliminated. However, in practice, to fully avoid the 0.95 p.u. violation, $V^{\text{min}}=0.955$ p.u. can be used.

\begin{table}[]
\centering
\caption{Voltage Violation by Controllers}
\label{table-voltage-violation}

\resizebox{\linewidth}{!}{\begin{tabular}{c|c|cccc>{\columncolor[gray]{0.9}}c>{\columncolor[gray]{0.9}}c}
\specialrule{.15em}{.075em}{.075em}
\multicolumn{2}{c|}{\textbf{Controller}}        & \textbf{RL-1}  & \textbf{RL-2}  & \textbf{RL-4} & \textbf{RL-6} & \textbf{NR-MPC} & \textbf{RC-MPC} \\
\specialrule{.15em}{.075em}{.075em}

\multirow{2}{*}{0}     & $\overline{\tau_{\text{vio}}}$      & 10.32 & 14.43 & 8.99 & 16.56 & 373.28 & 364.03 \\ \cline{2-8} 
                      & $\overline{\Tilde{V}_{\text{vio}}}$    & 0.9491 & 0.9492 & 0.9493 & 0.9489 & 0.9467 & 0.9467 \\ \hline
\multirow{2}{*}{5\%}     & $\overline{\tau_{\text{vio}}}$      & 19.81 & 17.12 & 10.42 & 21.55 & 358.10 & 330.15 \\ \cline{2-8} 
                      & $\overline{\Tilde{V}_{\text{vio}}}$    & 0.9490 & 0.9489 & 0.9487 & 0.9489 & 0.9468 & 0.9466 \\ \hline
\multirow{2}{*}{10\%}     & $\overline{\tau_{\text{vio}}}$      & 13.33 & 16.04 & 12.39 & 24.76 & 304.20 & 251.62 \\ \cline{2-8} 
                      & $\overline{\Tilde{V}_{\text{vio}}}$    & 0.9491 & 0.9491 & 0.9490 & 0.9488 & 0.9466 & 0.9459 \\ \hline
\multirow{2}{*}{15\%}     & $\overline{\tau_{\text{vio}}}$      & 13.33 & 8.18 & 14.10 & 11.07 & 287.41 & 204.66 \\ \cline{2-8} 
                      & $\overline{\Tilde{V}_{\text{vio}}}$    & 0.9491 & 0.9485 & 0.9491 & 0.9492 & 0.9465 & 0.9451 \\ \hline
\multirow{2}{*}{20\%}     & $\overline{\tau_{\text{vio}}}$      & 16.32 & 17.45 & 12.29 & 12.50 & 288.62 & 208.23 \\ \cline{2-8} 
                      & $\overline{\Tilde{V}_{\text{vio}}}$    & 0.9492 & 0.9490 & 0.9486 & 0.9490 & 0.9461 & 0.9446 \\ \hline
\multirow{2}{*}{25\%}     & $\overline{\tau_{\text{vio}}}$      & 14.49 & 12.59 & 21.67 & 19.03 & 299.62 & 222.12 \\ \cline{2-8} 
                      & $\overline{\Tilde{V}_{\text{vio}}}$    & 0.9491 & 0.9491 & 0.9491 & 0.9489 & 0.9459 & 0.9444 \\
\specialrule{.15em}{.075em}{.075em}
\end{tabular}}
\end{table}

\subsection{Comparison With Direct Learning}

To show the necessity of using CL in training, three state-of-the-art RL algorithms (one on-policy policy gradient method, one direct policy search and one off-policy Q-learning based method) are used to directly train controller for (\ref{eq-opt-formulation}) for comparison. These algorithms and their corresponding hyper-parameters we tuned are:

1) \textbf{PPO} \cite{schulman2017proximal}, a policy gradient method, trained with a learning rate in $\{10^{-5}, 10^{-4}, 10^{-3}\}$, on-policy training batch size in $\{3\times10^{4}, 10^{5}\}$ and entropy coefficient in $\{0.0, 0.005, 0.02\}$.

2) \textbf{ES-RL} \cite{salimans2017evolution}, a zero-order policy gradient estimation based method, trained with a learning rate in $\{5\times10^{-4}, 10^{-3}, 5\times10^{-3}\}$, objective function Gaussian smoothing standard deviation in $\{0.01, 0.02, 0.03, 0.04\}$. 

3) \textbf{Ape-X (DDPG)} \cite{horgan2018distributed}, a high throughput implementation of Q-learning based method, trained with a learning rate in $\{10^{-4}, 10^{-3}, 5\times10^{-3}\}$ and two Ornstein-Uhlenbeck noise schedule anneal after 6M/120M training steps.

Table \ref{table-coverged-reward} shows the comparison of converged reward between CL-based approach and directly learning (DL) approaches, and the DL results are based on the best grid searched hyper-parameters. It shows controllers trained by CL converges to higher reward than those trained by DL, even after adequate hyper-parameters tuning. As explained in Section \ref{subsec-dl-challenges}, by solving the simpler problem, the CL approach can warm-start RL with better initial network parameters in Stage II, and thus is less sensitive to the choice of hyper-parameters, making it possible to converge to a better policy.

\begin{table}[]
\centering
\caption{Converged Reward using Different RL Algorithms}
\label{table-coverged-reward}

\begin{tabular}{c|c|cccccc}
\specialrule{.15em}{.075em}{.075em}
\multicolumn{2}{c|}{$\epsilon_T$}        & 0.0  & 5\%  & 10\% & 15\% & 20\% & 25\% \\
\specialrule{.15em}{.075em}{.075em}

\rowcolor[gray]{0.9}\multicolumn{2}{c|}{Proposed CL} & 23.26 & 23.68 & 23.14 & 23.03 & 22.61 & 22.71 \\ \hline
\multirow{3}{*}{\rotatebox[origin=c]{90}{Direct}}     & PPO      & -5.21 & 0.24 & -26.74 & -19.47 & -3.40 & 9.09 \\ \cline{2-8} 
                      & ES-RL    & 18.07 & 18.38 & 18.01 & 17.52 & 17.63 & 16.86 \\ \cline{2-8} 
                      & Ape-X     & 13.45 & 14.01 & 12.66 & 10.18 & 5.70 & 13.30 \\
\specialrule{.15em}{.075em}{.075em}
\end{tabular}
\end{table}

\subsection{Larger Test System and Computational Requirement}  \label{subsec-larger-system}

\begin{figure}[]
\centering
\includegraphics[width=0.9\linewidth]{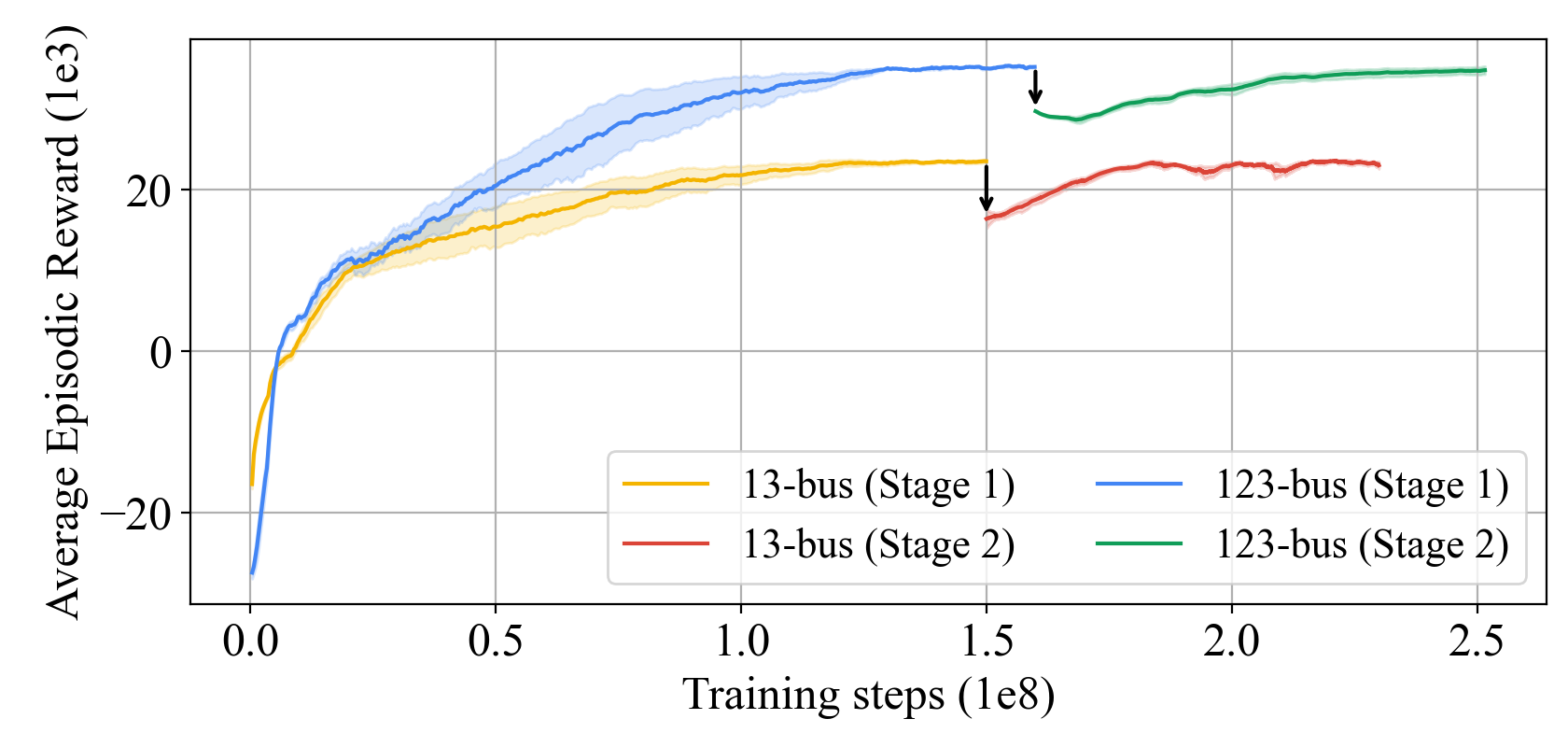}
\caption{Two stages learning curves comparison between 13-bus and 123-bus system. The learning curves are based on multiple runs with $\epsilon_T=10\%$ and $k=1$. Black arrow indicates the knowledge transfer between stages.}
\label{fig-computational-requirement}
\end{figure}

The proposed CL-based method is also tested in solving a CLR problem in a modified IEEE 123-bus test system. Table \ref{table-13-vs-123} compares the scale of two cases and shows the training steps ($TS_{\cdot}$) and wall time ($t^{\text{wall}}_{\cdot}$) required for convergence, and Fig. \ref{fig-computational-requirement} illustrates two sets of learning curves. According to the results, scaling from 13-bus system to 123-bus system does not increase the computational resources significantly. Unlike optimization based methods, in which the problem scale (e.g., number of variables) increases linearly with the number of buses, the RL problem's scale depends more on the state/action spaces, which are related to the numbers of DERs and loads. In this case, $|\mathcal{G}|$ and $|\mathcal{L}|$ does not increase remarkably and thus limits the increase of computational resources. In addition, in cases where multiple renewable generation share the same profile due to the proximity of location, the dimension of $\mathbf{s}_t$ might remain the same though $|\mathcal{R}|$ increases.

In our experiments, the back-propagation (BP) free ES-RL algorithm is leveraged for Stage I training to conduct a quick policy search for the simpler problem. It is scaled on 20 HPC computing nodes with a total of 719 parallel workers to accelerate learning \cite[Sec.2.1]{salimans2017evolution}. For Stage II, BP-based PPO algorithm is used for a more accurate policy search for (\ref{eq-opt-formulation}). Values of $t_{S2}^{\text{wall}}$ are based on PPO implementation using a single HPC node with 35 remote workers, and it can be further reduced using either high-throughput implementation or a global-local policy search method \cite{zhang2020grid}. 

\begin{table}[]
\centering
\caption{Scale Comparison for the Two Studied Cases}
\label{table-13-vs-123}

\begin{tabular}{c|cccccc}
\specialrule{.15em}{.075em}{.075em}

\textbf{System} & $|\mathcal{G}|$ & $|\mathcal{L}|$ & $TS_{S1}$ & $t_{S1}^{\text{wall}} (hour)$ & $TS_{S2}$ & $t_{S2}^{\text{wall}} (hour)$ \\ \hline
13-bus          & 4 & 15 & 1.5e8 & 0.27 & 8e7 & 16.77 \\ \hline
123-bus          & 6 & 30 & 1.6e8 & 0.31 & 9e7 & 19.91 \\ 
\specialrule{.15em}{.075em}{.075em}

\end{tabular}
\end{table}

\subsection{Additional RL benefits}

\begin{figure}[]
\centering
\includegraphics[width=0.8\linewidth]{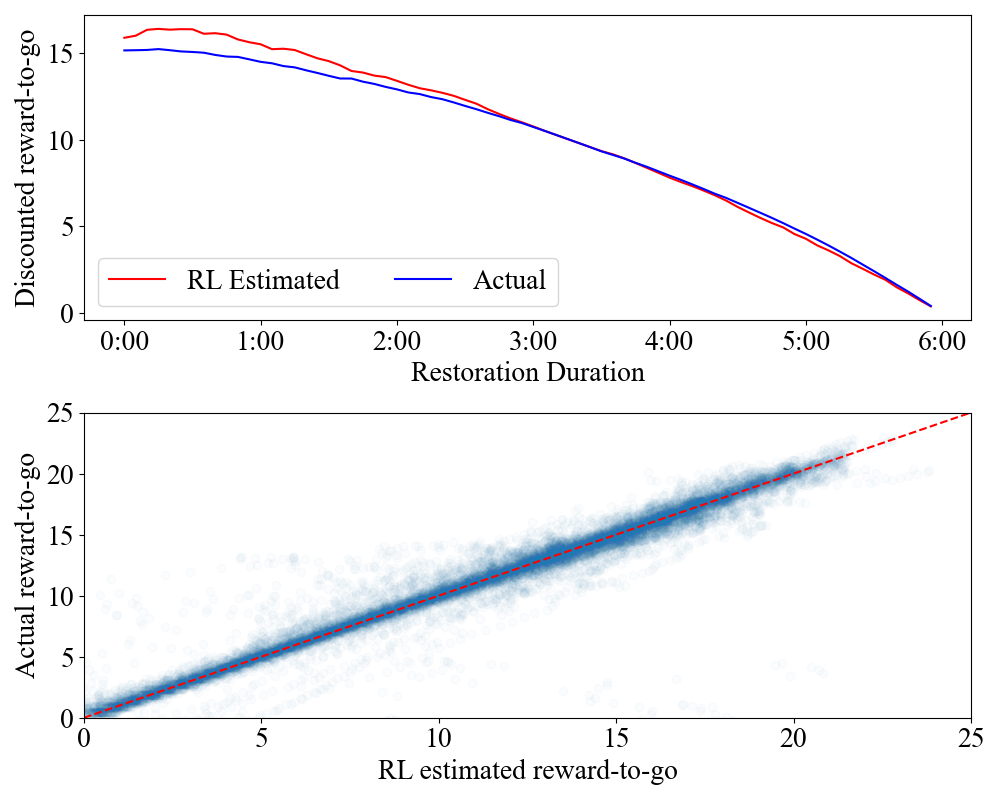}
\caption{\textit{Top}: Actual discounted reward-to-go ($G_t=\sum_{i=t}^{T} \gamma^{i-t} r_t$) vs. PPO value function estimated vale ($v_{\text{ppo}}(\mathbf{s}_t)$) in one example control episode. \textit{Bottom}: Comparison of single step $G_t$ vs. $v_{\text{ppo}}(\mathbf{s}_t)$ for 200 episodes (i.e., $200 \times T$ data points in the figure). Here, $\mathbf{s}_t$ is constructed with imperfect renewable forecasts with $\epsilon_T=10\%$.}
\label{fig-value-estimation}
\end{figure}

The critic network of Stage II trained PPO controller, $v_{\text{ppo}}$, can provide extra benefit, e.g., situation awareness, to system operators: at $t\in\mathcal{T}$, the expected discounted reward-to-go, i.e., $G_t=\sum_{i=t}^{T} \gamma^{i-t} r_t$ can be estimated by $v_{\text{ppo}}(\mathbf{s}_t)$. As shown in Fig. \ref{fig-value-estimation}, even though $\mathbf{s}_t$ contains inaccurate $\mathbf{\hat{p}}_{t}^r$, the estimated reward-to-go $v_{\text{ppo}}(\mathbf{s}_t)$ is, in most cases, a reliable indicator of $G_t$ calculated in retrospect. As a result, during the restoration, system operators have real-time estimation of the expectation of future restoration performance, and thus can use such information to facilitate other relevant decision-making.

\section{Discussion and Conclusion}

In this section, a methodological comparison between the RL approach and traditional optimization based methods (OBM) for CLR problems is provided; then, a conclusion is made and some future research directions are introduced.

\subsection{Discussion}

From the control performance perspective, even though being model-free and without optimality guarantee, RL based approach can achieve a good performance when compared with an OBM based method (Section \ref{subsubsec-load-restoration}). The exposure to a variety of scenarios during training also allows RL controllers to perform better than a deterministic MPC baseline that explicitly considers generation reserve when forecast errors are large. In addition to the impact of forecast errors, by integrating a more accurate and nonlinear model in the learning environment, using RL controllers can ameliorate performance deterioration caused by modeling error, e.g., power flow linearization (Section \ref{subsubsec-voltage-constraints}). 

Regarding constraints handling, unlike OBM, which naturally fold constraints into an optimization formulation, this is less straightforward in RL approaches. In this study, we introduced two techniques to take grid control operational constraints into consideration: 1) adding a penalty term in the objective function (Section \ref{subsec-opt-obj}), and 2) including an action preprocessor in the learning environment to map actions given by the policy network to the feasible region (Section \ref{subsubsec-learning-env}). Though being implicit, our experiments show that these techniques can effectively enforce constraints, making RL actions feasible.

Admittedly, compared with well-studied OBM-based approaches, RL-based methods still have limitations that requires further investigation. For example, considering network reconfiguration/switch operation is more challenging than the OBM counterparts: 1) advanced neural network structures, such as a graph neural network, are required to take in the current topological information at each control step; 2) a careful policy network outputs design is needed as the control actions become ``mixed integer''; and 3) a mechanism to handle non-stationary action space is needed since actions related to switches become inactive once the switch is closed and thus shrinks the action space. Due to these complexities, which we believe deserve a full paper to comprehensively investigate, we explained in [A.3] in Section \ref{sec-clr-problem} that network reconfiguration is deferred to future works, but we would like to provide the discussion here as an explanation.

\subsection{Conclusion}
In this study, we investigated leveraging deep RL to solve the CLR problem in distribution systems, using imperfect renewable forecasts for decision-making. To address the challenge posed by grid control problem complexity and non-convex RL policy search, we developed a two-stage learning curriculum to train an RL controller for the CLR problem. Using this approach, experiments show that the policy search can converge to a better policy than those learned directly with the original complex problem. The trained RL control policies, upon testing on unseen scenarios and compared with two MPC baselines, demonstrate proper optimal load restoration behavior and more robust performance in cases where renewable generation uncertainty is large. Through this example, the effectiveness of CL-based approach and the efficacy of RL controller for a grid control problem with uncertainty are manifested, and we hope this will inspire researchers to use RL to solve grid control problems with more complex nature. In future work, we will adapt the RL controller for more realistic scenarios, e.g., by considering the unfolding of the extreme events; and investigate whether RL can deliver a chance constrained policy for grid operation under uncertainty.

\appendices

\setcounter{equation}{0}
\setcounter{table}{0}
\renewcommand\theequation{A.\arabic{equation}}
\renewcommand\thetable{A.\Roman{table}}

\section{Formulation for RC-MPC} \label{appen-rc-mpc-formulation}

In RC-MPC, the objective is modified as:

\begin{equation} \label{eq-rc-mpc-obj-func}
r' := \sum_{t\in \mathcal{T}} (r_t^{\text{CLR}} + \vartheta_t - \phi[c\sum_{r\in\mathcal{R}} p_t^r - \sum_{g\in\mathcal{D}} p_t^{g, \text{res}}]^+),
\end{equation}
adding a third term to consider the penalty for failing to meet a rule based reserve requirement. In (\ref{eq-rc-mpc-obj-func}), $\mathcal{D}:=\{\mathcal{D}^{f} \cup \mathcal{D}^{s}\}$ collects all dispatchable resources. Parameter $\phi$ is the unit cost for reserve requirement violation, $p_t^{g, \text{res}} \in [0, \overline{p^{g}}]$ is the reserved generation for a dispatchable DER $g\in\mathcal{D}$, and $c$ is a configurable reserve requirement coefficient. Correspondingly, constraints for dispatchable DERs, i.e., $g \in \mathcal{D}^{f}$ and $\theta \in \mathcal{D}^{s}$, have the following modifications:

\begin{align}  \label{eq-rc-mpc-fuel-der-constraints}
\begin{split}
\underline{p^{g}} \leq p_t^{g} + p_t^{g, \text{res}} \leq \overline{p^{g}},\quad \\
    \alpha_t^g \in [\underline{\alpha^g}, \overline{\alpha^g}],\quad
    \sum_{t \in \mathcal{T}} (p_t^{g} + p_t^{g, \text{res}}) \cdot \tau \leq E^{g}
\end{split}
\end{align}

\begin{align}  \label{eq-rc-mpc-storage-constraints}
\begin{split}
-p^{\theta, \text{ch}} \leq p_t^{\theta}, \quad  p_t^{\theta} & + p_t^{\theta, \text{res}} \leq p^{\theta, \text{dis}}, \quad
S_{t+1}^{\theta} = S_{t}^{\theta} - \eta_t \cdot p_t^{\theta} \cdot \tau\\
\underline{S^{\theta}} \leq S_{t}^{\theta} \leq \overline{S^{\theta}} &, \quad
S_0^{\theta} = s_0, \quad
\alpha_t^{\theta} \in [\underline{\alpha^{\theta}}, \overline{\alpha^{\theta}}],
\end{split}
\end{align}

Regarding the reserve requirement, with a larger $c$, the RC-MPC becomes more conservative; and if $c=0$, it degenerates to NR-MPC. In this study, we increase the value of $c$ with the increment of error level, see Table \ref{table-c-values} for values used in Section \ref{sec-case-study}.

\begin{table}[]
\centering
\caption{Reserve Configuration under Different Error Levels}
\label{table-c-values}

\begin{tabular}{c|cccccc}
\specialrule{.15em}{.075em}{.075em}

$\epsilon_T$ & 0.0  & 5\%  & 10\% & 15\% & 20\% & 25\% \\ \hline
$c$                       & 10\% & 20\% & 40\% & 60\% & 75\% & 75\% \\ 
\specialrule{.15em}{.075em}{.075em}

\end{tabular}
\end{table}

As a result, the RC-MPC solves the following problem:

\begin{equation} \label{eq-rc-mpc-opt-formulation}
\begin{aligned}
& \underset{\mathbf{p}_t, \mathbf{q}_t, \mathbf{p}^\mathcal{G}_t, \bm{\alpha}^\mathcal{G}_t, \forall t \in \mathcal{T}}{\operatorname{maximize}} 
& & (\ref{eq-rc-mpc-obj-func}) \\
& \underset{\forall t \in \mathcal{T}}{\text{subject to}}
& & (\ref{eq-rc-mpc-fuel-der-constraints}) - (\ref{eq-rc-mpc-storage-constraints}), (\ref{eq-renewable-constraints}) - (\ref{eq-power-flow-constraints}).
\end{aligned}
\end{equation}

Note, when comparing with RL controller, the reserve penalty term in (\ref{eq-rc-mpc-obj-func}) is removed.

\setcounter{equation}{0}
\renewcommand\theequation{B.\arabic{equation}}

\section{Proof of Proposition 1} \label{appen-proof-proposition-1}

For actual renewable generation over a period of $T$ steps $\mathbf{p}^r = [p_0^r, p_{1}^r, ..., p_{T}^r]^\top$, there is:
\begin{equation}
    p_{k+1}^r = p_{k}^r + \Delta p_{k}^r,
\end{equation}
where $\Delta p_{k}^r$ reflects the generation variation between steps caused by the underlying dynamics, e.g., change of wind speed or cloud coverage for solar. Given the initial generation $p_0^r$, a forecaster predicts the sequence of $\Delta p_{k}^r$ for $k\in [0, T-1]$. However, with the forecast error $\iota_i$, the predicted generation becomes:
\begin{equation}
    \hat{p}_{k+1}^r = \hat{p}_{k}^r + \Delta p_{k}^r + \iota_{k} p^{r, \text{max}}.
\end{equation}

Suppose the single-step normalized forecast error $\iota_i \sim N(0, \left(\epsilon_T \sqrt{\frac{\pi}{2T}}\right)^2)$.
Note, the forecaster is assumed to be unbiased, meaning it tends to neither over-forecast nor under-forecast, so that $\iota_i$ is zero-mean. Now, the T-step normalized prediction error $|p_{T}^r - \hat{p}_{T}^r| / p^{r, \text{max}}=|\sum_{k=0}^{T-1} \iota_{k}|$ follows the \textit{half-normal distribution}, as it is the absolute value of a sum of normal random variables (which itself is normally distributed).

As $\sum_{k=0}^{T-1} \iota_{k} \sim N(0, T \left(\epsilon_T \sqrt{\frac{\pi}{2T}}\right)^2) = N(0, \left(\epsilon_T \sqrt{\frac{\pi}{2}}\right)^2)$, and $|p_{T}^r - \hat{p}_{T}^r| / p^{r, \text{max}}$ is half-normal, it follows that
\begin{equation}
    \mathbb{E}[|p_{T}^r - \hat{p}_{T}^r| / p^{r, \text{max}}] = \frac{\epsilon_T \sqrt{\frac{\pi}{2}}\sqrt{2}}{\sqrt{\pi}} = \epsilon_T.
\end{equation}

For the standard deviation, note the variance of the accumulation is $\left(\epsilon_T \sqrt{\frac{\pi}{2}}\right)^2$. Therefore the variance of the associated half-normal distribution is $\left(\epsilon_T \sqrt{\frac{\pi}{2}}\right)^2( 1 - \frac{2}{\pi} )$, and the standard deviation is $\epsilon_T\sqrt{\frac{\pi}{2} - 1}$.
\qed

\ifCLASSOPTIONcaptionsoff
  \newpage
\fi



%

\bibliographystyle{IEEEtran}
\bibliography{reference}




\end{document}